\documentclass[11pt,letterpaper]{article}

\usepackage{amsmath}
\usepackage{amsfonts}
\usepackage{amssymb}
\usepackage{stmaryrd}
\usepackage{array}
\usepackage[pdftex]{graphicx}
\usepackage{soul}
\usepackage{url}
\usepackage{multicol}
\usepackage{doi}   
\usepackage{cancel} 

\usepackage{bm}

\usepackage{booktabs}
\usepackage[dvipsnames]{xcolor}
\usepackage{tabularx} 
\usepackage{float}
\usepackage[margin=1in,letterpaper]{geometry} 
\usepackage[normalem]{ulem} 
\usepackage{setspace} 

\newtheorem{ass}{Assumption}

\newtheorem{defin}{Definition}

\newtheorem{lem}{Lemma}

\newtheorem{prop}{Proposition}

\begin{document}

\title{\textsc{Matching, Unanticipated Experiences, Divorce, Flirting, Rematching, Etc.}\thanks{We thank Kirill Rudov, Gaoji Hu, and Tyler Hoppenfels for helpful discussions and comments. Burkhard gratefully acknowledges financial support via ARO Contract W911NF2210282.}}

\author{Burkhard C. Schipper\thanks{Department of Economics, University of California, Davis. Email: bcschipper@ucdavis.edu} \and Tina Danting Zhang\thanks{Department of Economics, University of California, Davis. Email: tdzhang@ucdavis.edu}}

\date{May 12, 2025}
	
\maketitle
	
\begin{abstract} We study dynamic decentralized two-sided matching where players' preferences evolve due to unanticipated experiences. Stability requires no pairwise common belief in blocking, but unanticipated experiences can destabilize matchings. We show the existence of self-confirming outcomes that are stable and do not lead to unanticipated experiences. We propose a decentralized matching process that prioritizes mutual optimal blocking pairs with probability $1 - \varepsilon$ and picks any other optimal blocking pair with $\varepsilon$, representing market friction. Frictions are necessary for convergence to self-confirming stable outcomes even without unawareness. We extend our results by allowing communication and show convergence to flirt-proof self-confirming outcomes.   
\newline
\newline
\textbf{Keywords:} Decentralized matching, unawareness, transformative experiences, endogenous preferences, disclosure.  
\newline
\newline
\textbf{JEL-Classifications:} D83, C70.
\bigskip
\end{abstract}

\newpage

\section{Introduction} 

Many matching problems such as whom to marry, which school to choose, which profession to enter, where to take up a residency etc. involve transformative experiences (Paul, 2014) that change who we are, our beliefs, the things and issues we care about, and our preferences. These preference changes have implications for the stability of matchings. For instance, in the US marriage market, 43\% of ever-married couples were divorced or widowed, and 23\% of married couples are remarried couples (Livingstone, 2014). During marriage, spouses may become aware of intolerable attributes of the partner that they were previously oblivious to it. Or they may experience events in the family and workplace such as addiction, unemployment, domestic violence, mental health issues etc., some of which can only be fully grasped once experienced. For instance, marriage often involves parenting, itself a transformative experience, with profound changes in preferences as spouses become parents. Svar and Verner (2008) found a negative causal impact of children on relationship duration in Denmark. 

Importantly, the transformative experiences, consequent preference changes, and implications for the matchings cannot be completely anticipated and comprehended before they are experienced. Again taking the marriage market as an example, Baker and Emery (1993) report that the median response to the question asking a non-representative sample of marriage license applicants to estimate the fraction of US couples who marry will divorce was 50\%, while the median response assessing the likelihood that they personally would divorce was 0\%. Maher (2003) finds similar numbers of 52\% and 10\% respectively, in non-representative samples of the general population, and 48\% and 17\%, respectively, for law students. Svarer and Verner (2008) find that the first child but not later children are associated with the dissolution of a relationship, pointing to the causal effect of unanticipated rather than anticipated events surrounding parenting on divorce.\footnote{There is evidence from other matching markets as well. For example in the labor market, on average among baby boomers, men held 12.8 jobs and women held 12.5 jobs from ages 18 to 56 (U.S. Department of Labor, 2023). Using data from employees in a financial institution, Holtom et al. (2017) state that employees report a substantial number of unanticipated shocks, both personal and organizational, and that only unanticipated shocks were significant predictors of staff turnover while none of the anticipated shocks were.} 

How should we understand the dynamics and stability of matching markets with transformative experiences and unanticipated preferences changes? The elegant standard matching model of Gale and Shapley (1962) (see Roth and Sotomayor, 1990) gives little guidance as it is static. Models of decentralized matching processes (e.g., Knuth, 1976, Roth and Vande Vate, 1990, Ackermann et al., 2008, Rudov, 2024) focus on the path to stability without considering that experiences in matchings can lead to preference changes. With a few exceptions (e.g., Lazarova and Dimitrov, 2017, Chen and Hu, 2020), these approaches also lack incomplete information about preferences. Recent interesting models of matching under incomplete information allow mostly only for one-sided incomplete information, are static, and unrealistically assume that agents can anticipate all relevant future experiences (e.g., Liu, 2020, Bikhchandani, 2017, Liu et al., 2014, Pomatto, 2022, Forges, 2004). Because of imperfect information, these models require also sophisticated solution concepts that make use of information revealed from absence of blocking and from counterfactual blocking, which add to the complexity of applying and analyzing matching under incomplete information. In order to capture transformative experience with unanticipated preference changes, we develop dynamic matching games under unawareness but perfect information in Section~\ref{sec:unawareness}. Unawareness refers to the lack of conception rather than the lack of information and thus provides us with a robust notion of being ``unanticipated''. We make use of unawareness structures introduced by Heifetz, Meier, and Schipper (2006, 2008, 2013) to capture asymmetric unawareness among players but simplify them to the case of perfect information with respect to what players are aware of. These structures are complemented with finite state machines that model the change of awareness in experienced matchings and the resulting preference changes.\footnote{Our class of dynamic matching games with unawareness have the flavor of stochastic games, another invention by Shapley (1953), in non-cooperative game theory except that ours are cooperative games of matching, state transitions are deterministic, and we allow for asymmetric unawareness.} Rather than allowing any arbitrary preference changes, we only consider preference changes due to players becoming aware of events during matchings.

We adapt the notion of stability to asymmetric unawareness by requiring absence of pairwise common (point)-belief in blocking. This reflects the idea that individual willingness to block, even when it coincidentally occurs with the blocking partner, by itself may not necessarily result in a new match but it takes the \emph{agreement} of a pair to block. Such a notion of stability is consistent with Wilson's (1978) notion of the coarse core that he introduced for exchange economies with asymmetric information and that has been extended to coalitional transferable utility games with unawareness by Bryan, Ryall, and Schipper (2022). Because experiences in a matching, even in a stable matching, can lead to changes in awareness and thus preferences changes, stability of the matching is not enough as a solution concept in the dynamic matching game with unawareness. We also need that awareness and thus preferences do not change in the stable matching. In other words, the state determining awareness and thus preferences must be absorbing (w.r.t. the finite state machine modeling the awareness dynamics based on experiences in matchings). A self-confirming outcome is a pair of a matching and state such that the matching is stable w.r.t. awareness and thus preferences at the state, and the state is absorbing given the matching (see Section~\ref{sec:selfconfirming}).\footnote{Our terminology is inspired by self-confirming equilibrium in non-cooperative game theory, which are outcomes in which players maximize expected utility w.r.t. beliefs consistent with their observations and these observations are generated by the play in equilibrium (see for instance Fudenberg and Levine (1993) and Battigalli and Guaitoli (1997) for games without unawareness and Schipper (2021) for games with unawareness).} 

To model the process of matching and re-matching, we first revisit the convergence of decentralized matching processes to stability for fixed preferences without unawareness in Section~\ref{sec: revisited}. Knuth (1976) showed that a deterministic process of satisfying blocking pairs may lead to cycles. However, his example is unnatural in the sense that when there are multiple blocking pairs at a stage, he satisfies a blocking pair that is not mutually optimal. A blocking pair is optimal for a player if it is the best blocking pair for the player. It is mutually optimal if it is the best blocking pair for both players. In markets without frictions, we would expect that mutually optimal blocking pairs are satisfied whenever they exist. The example by Knuth (1976) leaves open the conjecture that a process of satisfying mutually optimal blocking pairs, when they exist, leads to the stability. We show with a new example that this is not the case. This implies that for the process of randomly picking blocking pairs to converge to stability as in Roth and Vande Vate (1990), it is necessary that strict positive probability is assigned to satisfying blocking pairs that are \emph{not} mutually optimal blocking pairs. This yields an important corollary: Frictions are necessary for decentralized matching processes to converge to stability. This is contrary to the intuition in economics that markets function best in the absence of frictions. 

Armed with the insights from our study of decentralized matching processes with full awareness, we introduce in Section~\ref{sec:selfconfirming} a decentralized matching process that starts with any unstable matching, satisfies a randomly picked mutually optimal blocking pair with probability $1 - \varepsilon$ if it exists, and satisfies an optimal blocking pair otherwise. W.r.t. states (and hence awareness), the dynamics follows the finite state machine mentioned above. We show that this decentralized process converges to a self-confirming outcome. Moreover, we demonstrate by example that in such a self-confirming outcome, players may remain unaware. 

In real life, awareness may not just be raised via experiences in matches but also through communication. In Section~\ref{sec:flirting}, we extend the model to communication by allowing players to raise the other players' awareness with the intend to create pairwise common belief in blocking (i.e., ``flirting''). Formally, this is modeled with another finite state machine. Communication can have two kinds of effects in our model: First, because it potentially raises the other players' awareness, it can change their preferences and thus create blocking pairs. Second, it may raise awareness in such a way to augment blocking pairs with pairwise common belief in blocking. In both cases, such communication may invite further communication, leading to further changes of awareness, beliefs, and preferences etc. A flirt-proof stable outcome consists of a matching and state such that communication does not change the state given the matching and the matching is stable given the state.\footnote{Flirt-proof stability is reminiscent of extensions of Wilson's (1978) core concepts for exchange economies under asymmetric information, such as the fine core, that allow for various forms of information revelation; see Forges and Serrano (2013) for a survey.} When flirt-proof stable outcome is absorbing (w.r.t. the finite state state machine modeling changes in awareness given the experience in matchings), then we call it a flirt-proof self-confirming outcome. We show convergence of our decentralized matching process with communication to flirt-proof self-confirming outcome and that it refines the set of self-confirming outcomes. 

It is often argued that divorce improves well-being as it avoids the need for estranged spouses to suffer through a marriage. Longitudinal studies of divorce do not find clear-cut systematic improvement of well-being or mental health (Spanier and Furstenberg, 1982, Lucas, 2005, Symoens et al., 2013). While divorce allows one to escape a match, it is far from clear that the subsequent dynamic rematching process leads to a better outcome. One might hypothesize that the person initiating a divorce, the divorcer, should be better offer w.r.t. her changed preferences while the person that is divorced, the divorcee, might become worse off as he/she loses her/his partner that was matched in the prior stable matching. While earlier studies (Spanier and Furstenberg, 1982) find no evidence for improved well-being of the divorcer, more recent studies (Symoens et al., 2013) find a significant positive effect of being the one initiating the divorce. We show in Section~\ref{sec:divorce} that the welfare of divorcers can go in any direction even for the same matching game and same initial conditions. That is, divorcers may end up better, equivalent, or worse off w.r.t. their preferences after an awareness change. The same we show for the divorcee. So in terms of welfare effects of divorce, anything goes. 

In the final Section~\ref{sec:discussion}, we discuss the effect of infidelity. We also discuss how players can be confused about other players in self-confirming outcomes leading to (stable) awareness of unawareness (Schipper, 2024). Finally, we discuss the related literature.

\section{Decentralized Matching (without Unawareness) Revisited\label{sec: revisited}} 

In this section, we revisit standard decentralized matching without unawareness in order to show that arbitrarily small frictions are necessary for stable outcomes to emerge in decentralized random matching. This will motivate our decentralized random matching process that we use in later sections on matching with unawareness. 

Consider a standard two-sided marriage matching market with non-transferable utility (and without unawareness) $\langle M, W, (\succ_m)_{m \in M}, (\succ_w)_{w \cup W} \rangle$ with a nonempty finite set of men $M$, a nonempty finite set of women $W$, and for each man $m \in M$, $\succ_m$ is a strict preference relation over $W \cup \{m\}$ while for each woman $w \in W$, $\succ_w$ is a strict preference relation over $M \cup \{w\}$. 

A \emph{matching} is a one-to-one function $\mu: M \cup W \longrightarrow M \cup W$ such that $\mu(m) = w$ if and only if $\mu(w) = m$. If for $i \in M \cup W$, $\mu(i) = i$, then $i$ is unmatched. Given a matching $\mu$, a man $m$ and a woman $w$ form a \emph{blocking pair} if $\mu(m) \neq w$, $m \succ_w \mu(w)$, and $w \succ_m \mu(m)$. That is, $m$ and $w$ form a blocking pair if they are not matched to each other given $\mu$ and they prefer each other over their current match, respectively. A matching $\mu$ is individually rational if $\mu(i) \succeq_i i$ for all $i$. A matching $\mu$ is \emph{stable} if it is individually rational and there is no blocking pair.

Consider an initial matching $\mu$ and a decentralized process of successively satisfying blocking pairs. A blocking pair is \emph{satisfied} if they leave their current match and are match to each other, leading to a new matching. When does a process satisfying blocking pairs lead to a stable matching?

\subsection{Knuth's Cycle}

Knuth (1976) showed with the help of an example that the decentralized process of satisfying blocking pairs may not lead to a stable matching. We briefly present his cyclic example and discusses its weaknesses.\\

\noindent \textbf{Example 1 (Knuth, 1976)} Consider $M = \{m_1, m_2, m_3\}$ and $W = \{w_1, w_2, w_3\}$ with strict preferences given by the following rank order lists (from most preferred to least preferred, respectively):
$$\begin{array}{ll}
\succ_{m_1}: & w_2, w_1, w_3\\
\succ_{m_2}: & w_1, w_3, w_2\\
\succ_{m_3}: & w_1, w_2, w_3\\
\succ_{w_1}: & m_1, m_3, m_2\\
\succ_{w_2}: & m_3, m_1, m_2\\
\succ_{w_3}: & m_1, m_3, m_2
\end{array}$$

\begin{figure}[h]
    \centering
    \includegraphics[width=1\textwidth]{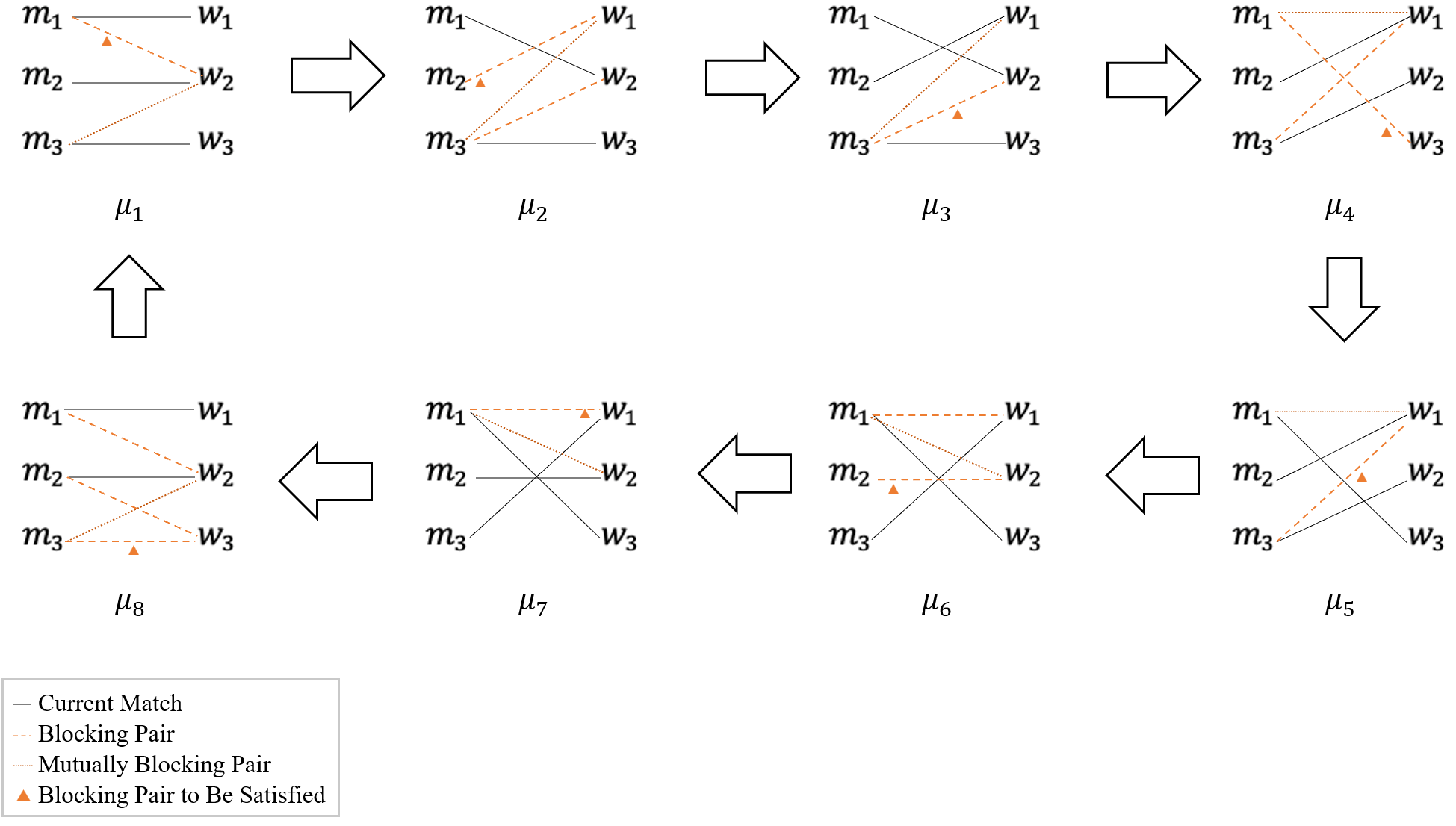}
    \caption{Knuth's Cycle}
    \label{fig:Knuth}
\end{figure}

Knuth's (1976) example involves a cycle of eight matchings.\footnote{We use the exposition of Knuth's example from Roth and Vande Vate (1990), as it uses the prevalent notations while the original exposition does not. } Let $\mu_1 = \begin{pmatrix}
m_1 & m_2 & m_3 \\
w_1 & w_2 & w_3
\end{pmatrix}$ be the first matching. That is, in matching $\mu_1$, man $m_1$ is matched to woman $w_1$, man $m_2$ is matched to $w_2$, etc. This matching is not stable because there are blocking pairs $(m_1, w_2)$ and $(m_3, w_2)$. Satisfying $(m_1, w_2)$ leaves their original matched partners, $w_1$ and $m_2$, unmatched and leads a new matching $\mu_2=\begin{pmatrix}
m_1 & m_2 & m_3 & w_1 \\
w_2 & m_2 & w_3 & w_1
\end{pmatrix}$. Subsequent matchings of the process are shown in the bipartite graphs in Figure~\ref{fig:Knuth}, where each matching is a bipartite graph, a match is represented with a solid line, and all blocking pairs are indicated with either dashed or dotted lines. The solid triangles represent the blocking pairs that are satisfied in order to reach the next matching. \\

As shown in Figure~\ref{fig:Knuth}, when we satisfy blocking pairs in the sequence indicated by solid triangles, the process loops back to the initial matching after eight rounds. Thus, the example shows that deterministic decentralized matching processes can lead to cycles that prevent the emergence of a stable matching. Deterministic refers to the fact that when there are multiple blocking pairs, we do not choose one randomly to satisfy but pick a particular with probability one. \hfill $\Box$\\

At a second glance, the Knuth's example is not fully convincing because it involves satisfying a sequence of rather unnatural blocking pairs. For example, in order to create Knuth's cycle, at matching $\mu_1$ the blocking pair $(m_1, w_2)$ needs to be satisfied instead of blocking pair $(m_3, w_2)$. Both blocking pairs involve woman $w_2$. Given that potentially both men, $m_3$ and $m_2$ compete for being matched to her, it is much more natural to satisfy blocking pair $(m_3, w_2)$ because $w_2$ prefers $m_3$ over $m_1$. If $w_2$ is asked to rematch, she would surely pick $m_3$ instead of $m_1$.

In order to formalize our observation, we follow Bennett (1994) in saying that pair $(i, j)$ is the \emph{optimal blocking pair} for $i$ at matching $\mu$ if $j$ is $i$'s most preferred individual among the set of individuals who forms a blocking pair with agent $i$ at the matching $\mu$. Furthermore, we call $(i, j)$ a \textit{mutually optimal blocking pair} at $\mu$ if it is an optimal blocking pair for both $i$ and $j$ at $\mu$. The mutually optimal blocking pairs are represented in Figure~\ref{fig:Knuth} by the dotted lines.

Knuth's cycle avoids choosing the optimal blocking pairs whenever they exist, which is in every round. We think this is very unnatural because in a frictionless marriage market optimal blocking pairs should be able ``to meet'' and consequently form a match. If in Knuth's example a mutually optimal blocking pair is chosen at any round, the process is guaranteed to get out of the cycle. For example, if we satisfy the mutually optimal blocking pair $(m_3, w_2)$ at $\mu_1$ instead of $(m_1, w_2)$, we reach the matching $\mu_2' = \begin{pmatrix}
m_1 & m_2 & m_3 & w_3 \\
w_1  & m_2   & w_2 & w_3
\end{pmatrix}$, which has only one blocking pair, $(m_2, w_3)$. Satisfying $(m_2, w_3)$ leads to a stable matching $\mu_3' = \begin{pmatrix}
m_1 & m_2 & m_3  \\
w_1 & w_3  & w_2
\end{pmatrix}$. Thus, while Knuth's example demonstrates that deterministic decentralized matching may lead to a cycle, it leaves open the possibility that deterministic decentralized matching in which more naturally at each step a mutual optimal blocking pair is satisfied, whenever it exist, leads to a stable outcome. Unfortunately, we show in the next section that this is not the case.

\subsection{A Cycle with Mutually Optimal Blocking Pairs\label{section:Zhang}}

In this section, we improve upon Knuth's cycle. We present a new example with a cycle that involves satisfying mutually optimal blocking pairs. At every stage, there exists a \emph{unique} mutually optimal blocking pair. When satisfying the unique mutually optimal blocking pair at every stage, the process ends up in a cycle. Compared to Knuth's cycle, we need now four participants on each side of the market.\\ 

\noindent \textbf{Example 2 } Consider a marriage market with four men $M = \{m_1, m_2, m_3, m_4\}$ and four women $W = \{w_1, w_2, w_3, w_4\}$, with the strict preference profiles, respectively, given by the rank order lists:
$$\begin{array}{ll}
\succ_{m_1}: & w_2, w_4, w_3, w_1\\
\succ_{m_2}: & w_4, w_2, w_1, w_3\\
\succ_{m_3}: & w_1, w_3, w_2, w_4\\
\succ_{m_4}: & w_3, w_1, w_4, w_2\\
\succ_{w_1}: & m_1, m_2, m_4, m_3\\
\succ_{w_2}: & m_4, m_3, m_2, m_1\\
\succ_{w_3}: & m_2, m_1, m_3, m_4\\
\succ_{w_4}: & m_3, m_4, m_1, m_2
\end{array}$$

\begin{figure}[h]
    \centering
    \includegraphics[width=1\textwidth]{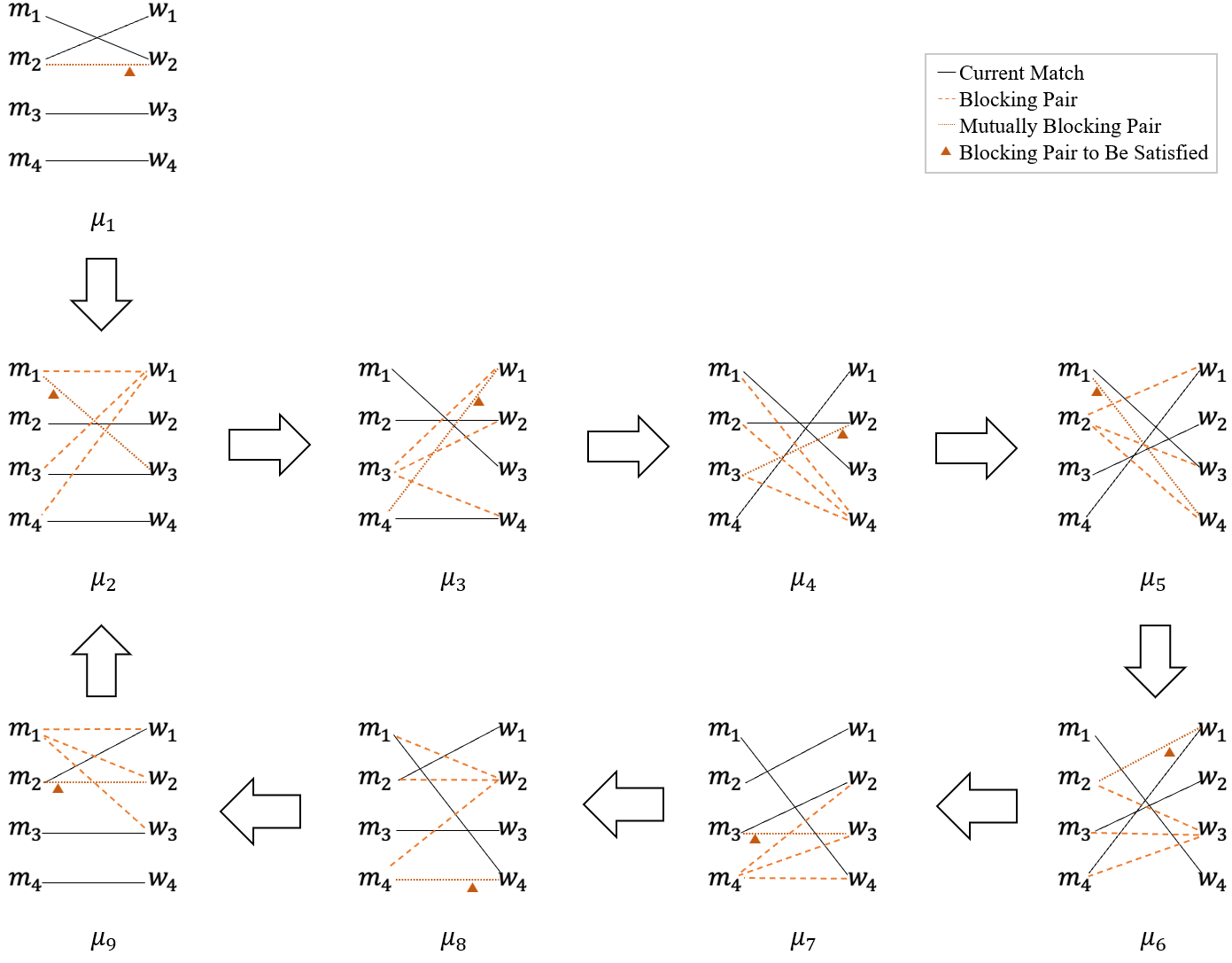}
    \caption{Cycle with Unique Mutually Optimal Blocking Pairs Only}
    \label{fig:Zhang}
\end{figure}

Let the first matching be given by $\mu_1 = \begin{pmatrix}
m_1 & m_2 & m_3 & m_4  \\
w_2 & w_1 & w_3 & w_4
\end{pmatrix}$. It has exactly one blocking pair $(m_2, w_2)$, which is also a mutually optimal blocking pair. Satisfying this blocking yields the second matching $\mu_2 = \begin{pmatrix}
m_1 & m_2 & m_3 & m_4  & w_1\\
 m_1 & w_2 & w_3 & w_4 & w_1
\end{pmatrix}$. Subsequent steps are shown in the sequence of bipartite graphs (each representing a matching) in Figure~\ref{fig:Zhang}.
After nine rounds of satisfying the unique mutually optimal blocking pairs, respectively, we reach the matching $\mu_{10} = \mu_2$, completing the cycle. This example improves Knuth's example by demonstrating that even when satisfying unique mutually optimal blocking pairs at each step of the process, we can get into a cycle. \hfill $\Box$\\

More generally, we conclude: 

\begin{prop} There does not exist a decentralized process of satisfying blocking pairs that always chooses mutually optimal blocking pairs when they exist and always reaches a stable matching.
\end{prop}

Kirill Rudov kindly informed us he has independently shown the same observation Rudov (2024, Proposition 4 in the online appendix). His counterexample makes use of a five-by-five market while we our counterexample involves just a four-by-four market.

\subsection{Random Paths to Stability}

Prior section showed that we need to go beyond deterministic processes in order to guarantee stable outcomes in decentralized matching markets. Roth and Vande Vate (1990) showed that we can reach stable outcomes when a process that randomly picks among blocking pairs. In particular, they showed that starting from an arbitrary matching $\mu$, there exist a finite sequence of matchings $\mu_1, ..., \mu_k$ such that $\mu = \mu_1$ and $\mu_k$ is stable, and for $\mu_i$, $i = 1, ..., k - 1$, there is a blocking pair $(m_i, w_i)$ that if satisfied yields matching $\mu_{i+1}$. As a corollary, they show that the process of satisfying randomly chosen blocking pairs will converge to a stable matching with probability 1. 

Our example in Section~\ref{section:Zhang} shows that the random process of Roth and Vande Vate does not always choose a mutually optimal blocking pair even when they exist. Thus, we conclude that the result by Roth and Vande Vate (1990) cannot be strengthened to prioritizing \emph{mutually optimal} blocking pairs (if they exist). More formally:

\begin{prop} The process of satisfying randomly chosen blocking pairs must put positive probability on sub-optimal blocking pairs (where at least one of the involved agent prefers another blocking pair) in order to converge to a stable matching. 
\end{prop}

This observation has a profound economic corollary: \emph{Frictions that prevent the satisfaction of mutually optimal blocking pairs are necessary for any decentralized matching processes to reach stability.} This observation runs counter to standard economic wisdom that frictions hamper the functioning of decentralized markets. In contrast, for decentralized matching markets, small frictions are necessary for reaching stable matchings. In a separate online-appendix, we present another application of our cyclic example to entry in matching markets.

\subsection{Arbitrarily Small Frictions are Enough}

While we just have shown that frictions are necessary to reach stable outcomes in decentralized random matching, we will now show that we can make these frictions arbitrarily small. 

We define the unperturbed process as follows: Starting from an arbitrary matching, if this matching is stable, no change occurs; if mutually optimal blocking pair(s) exist(s), satisfy one of them with equal probability; otherwise, satisfy one of the optimal blocking pair(s) with equal probability. Note that any unstable matching will have at least one optimal blocking pair. Apply the same rules to the next matching.

To introduce frictions, consider a perturbed process defined as follows: Starting from an arbitrary matching, if this matching is stable, no change occurs; if mutually optimal blocking pair(s) exist(s), with probability $1 - \varepsilon$ select randomly one mutually optimal blocking pair and satisfy it, and with probability $\varepsilon$ satisfy a randomly selected optimal blocking pair that is not mutually optimal; otherwise, satisfy a randomly selected optimal blocking pair. Apply the same rules to the next matching.

For the unperturbed process, both stable matchings and cycles formed by satisfying unique mutually optimal blocking pairs are absorbing sets. However, we argue that for an arbitrarily small $\varepsilon > 0$, which captures frictions that prevent the matching of mutually optimal blocking pairs, the perturbed process converges to a stable matching in finite time with probability 1. Given a marriage market $\langle M, W, (\succ_i)_{i \in M \cup W}\rangle$, cycle(s) formed by satisfying unique mutually optimal blocking pairs at each step may or may not exist. If there is no such cycle, our claim is trivially true. If there exists such a cycle, then conditional on entering such a cycle, the probability that the process stays in this cycle for exactly $k$ period is $(1 - \varepsilon)^{k-1}\varepsilon$. Hence the probability that the process stays in this cycle forever is $\lim_{k \rightarrow \infty} (1 - \varepsilon)^{k-1} \varepsilon = 0$, which means that the process leaves the cycle in finite time with probability 1. After leaving the cycle, the process might come back to this cycle or enter another cycle. However, the same argument applies, and the process leaves the cycle again in finite time with probability 1. As soon as the process catches one of Ackermann et al. (2008)'s optimal blocking pairs, it is trapped in a basin of attraction of a stable matching. Note that by the $\varepsilon$-events, any optimal blocking pair has a strict positive probability. Thus, we can make use of the result by Ackermann et al. (2008) according to which for any unstable matching, there exist a finite sequence of satisfying optimal blocking pairs that leads to a stable matching. The process may potentially spend long periods in the cycle(s), but only the stable matchings are absorbing  

The discussion so far motivates our decentralized process used in the following sections. We use a process of randomly chosen blocking pairs that prioritizes mutual optimal blocking pairs, whenever they exist, but at each step assigns arbitrarily small but non-zero probability to satisfying just optimal blocking pairs.

\section{Matching under Unawareness\label{sec:unawareness}}

We continue to consider a two-sided marriage matching market with non-transferable utility. However, different from the previous section we now allow for asymmetric awareness among players. The preference rankings of players can now depend on their awareness. We model asymmetric awareness using a simplified version of unawareness structures by Heifetz et al. (2006, 2013). Consider a finite lattice of disjoint finite state spaces $(\mathcal{S}, \trianglerighteq)$. Denote by $\check{S} = \bigvee_{S \in \mathcal{S}} S$ the join of the lattice. For any spaces $S, S' \in \mathcal{S}$ with $S' \trianglerighteq S$, there is a surjective projection $r^{S'}_{S}: S' \longrightarrow S$. Projections commute, i.e., for any $S, S', S'' \in \mathcal{S}$ with $S'' \trianglerighteq S' \trianglerighteq S$ we have $r^{S''}_{S} = r^{S'}_S \circ r^{S''}_{S'}$. Moreover, for any $S \in \mathcal{S}$, $r^{S}_S$ is the identity on $S$. Let $\Omega := \bigcup_{S \in \mathcal{S}} S$. We sometimes use $\omega_S$ to denote the projection of $\omega$ to space $S$. We also write $S_{\omega}$ for the space that contains $\omega$. 

For any $S \in \mathcal{S}$ and $D \subseteq S$, denote by $D^{\uparrow} := \bigcup_{S' \trianglerighteq S} (r^{S'}_{S})^{-1}(D)$. An \emph{event} $E \subseteq \Omega$ is defined by a base-space $S \in \mathcal{S}$ and a base $D \subseteq S$ such that $E := D^{\uparrow}$. Denote by $S(E)$ the base-space of event $E$. For any $S \in \mathcal{S}$, denote by $\Sigma(S)$ the set of events with base-space $S$ and by $\Sigma$ the set of all events.  

Awareness affects the preferences of players. To make it explicit, we let preferences of players and their (point-)beliefs depend on states. To this end, we introduce for each player a point-belief type mapping $t_i: \Omega \longrightarrow \Omega$ such that:
\begin{itemize}
\item[(i)] For any $S \in \mathcal{S}$, $\omega \in S$ implies $t_i(\omega) = r_{S'}^{S}(\omega)$ for some $S' \trianglelefteq S$. 
\item[(ii)] For any $S, S', S'' \in \mathcal{S}$ with $S'' \trianglerighteq S' \trianglerighteq S$, $\omega \in S''$, $t_i(\omega) \in S'$ implies $t_i(\omega_S) = r_{S}^{S'}(t_i(\omega))$. 
\item[(iii)] For any $S, S', S'' \in \mathcal{S}$ with $S'' \trianglerighteq S' \trianglerighteq S$, $\omega \in S''$ and $t_i(\omega_{S'}) \in S$ implies $S_{t_i(\omega)} \trianglerighteq S$.  
\end{itemize}   

Property (i) means that a player at a state cannot be aware of more than what is described by that state. Properties (ii) and (iii) are consistency conditions on how awareness is related across states. These properties specialize the properties of type mappings in Heifetz, Meier, and Schipper (2013) to our case of point-belief type mappings. That is, the structure so far is a special case of unawareness structures  in Heifetz, Meier, and Schipper (2013). Here we only allow for point-beliefs (rather than non-degenerate beliefs) and unawareness. We refer to $S_{t_i(\omega)}$ as player $i$'s awareness level at state $\omega$. Player $i$ with point-belief $t_i(\omega)$ can reason over all states in $S_{t_i(\omega)}$ and in spaces $S \trianglelefteq S_{t_i(\omega)}$. In particular, such a player is aware of all events with a base-space less expressive than $S_{t_i(\omega)}$. 

For simplicity, we restrict unawareness structures by ruling out redundancies.
\begin{ass}[No Redundancies]\label{ass: redundancies} For any $S \in \mathcal{S}$ and $\omega, \omega' \in S$ with $\omega \neq \omega'$, there exist $i \in M \cup W$ such that $S_{t_i(\omega)} \neq S_{t_i(\omega')}$.
\end{ass}
That is, different states imply that some player's awareness must differ.  

We also impose a richness condition according to which any combination of awareness among players is feasible. 

\begin{ass}[Richness]\label{ass: richness} For any profile of spaces $(S_i)_{i \in M \cup W} \in \mathcal{S}^{|M \cup W|}$, there exist $\omega \in \check{S}$ such that for any $i \in M \cup W$, $S_{t_i(\omega)} = S_i$.
\end{ass} 

\begin{figure}[h]
    \centering
    \includegraphics[width=0.8\textwidth]{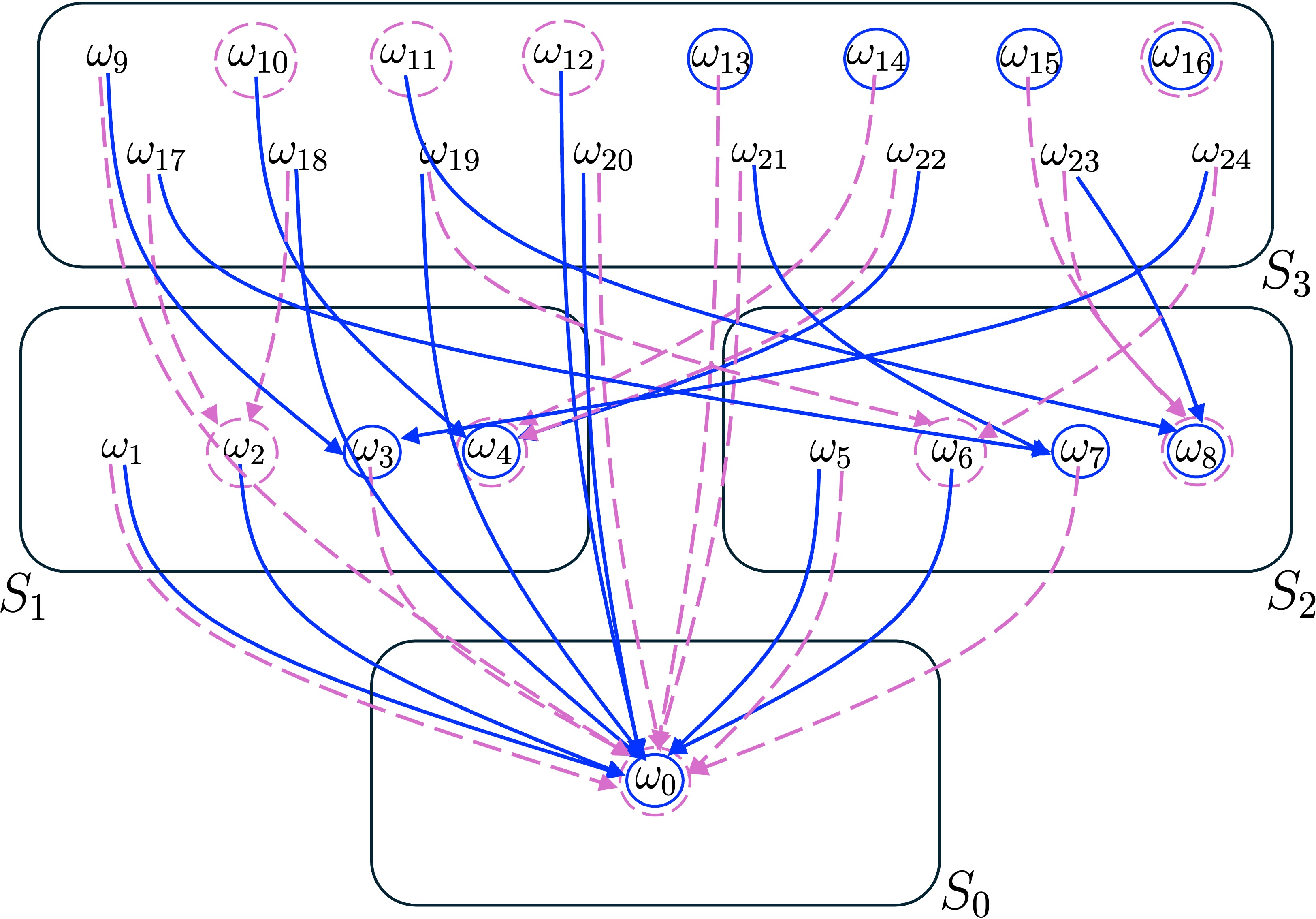}
    \caption{Illustration of an Unawareness Structure with Two Players and Two Characteristics}
    \label{fig:unawareness_structure}
\end{figure}
We illustrate an unawareness structure with point-beliefs for two players satisfying Assumptions~\ref{ass: redundancies} and~\ref{ass: richness} in Figure~\ref{fig:unawareness_structure}. There are two characteristics of players or experiences with players of which players may or may not be unaware. Thus, we have four spaces. The left space $S_1$ models the situation in which only the first characteristic/experience is expressible, while the right space $S_2$ models the situations when only the second characteristic/experience is expressible. On the upmost space $S_3$ both characteristics/experiences are expressible. In the lowest space $S_0$ none of the characteristics/experiences are expressible. States model combinations of awareness of both agents in a consistent way as described by the type mapping and the conditions imposed on the type mapping. The type mapping of one player (i.e., the man) is is indicated with blue solid lines while the one of the other player is indicated with pink dash lines (i.e., the woman). For instance, at state $\omega_{10}$ the woman is aware of both characteristics/experiences, because her type mapping maps $\omega_{10}$ to itself, while the man is only aware of the first characteristic/experience, because his type mapping maps $\omega_{10}$ to $\omega_4$ in $S_{1}$. At $\omega_4$ both players are aware of the first characteristic/experience. Thus, at $\omega_{10}$ the man point-believes that the woman is aware of the first characteristics/experiences, which is consistent with his awareness at $\omega_{10}$ since he can only envision the first characteristic/experience but not the second at $\omega_{10}$. More interesting is for instance state $\omega_{24}$. At that state, the man is aware of the first characteristic/experience only while the woman is only aware of the second characteristics/experiences only because their type mappings map to $\omega_3$ and $\omega_6$, respectively. At $\omega_3$, the man correctly believes that the woman is unaware of characteristic/experience 1 while at $\omega_6$ the woman correctly believes that the man is unaware of characteristic/experience 2. This is indicated by the type mappings that map the respective states to $\omega_0$. We omit the projections in order not to clutter the figure further, but they should be clear. For instance, states $\omega_{13}, \omega_{14}$ and $\omega_{21}$ all project to $\omega_7$. Assumptions~\ref{ass: redundancies} and~\ref{ass: richness} are easy to verify for instance in spaces $S_1$ and $S_2$ (albeit less obvious in $S_3$). Note that any unawareness structure for correct point-beliefs of two players and two characteristics/experiences satisfying Assumptions~\ref{ass: redundancies} and~\ref{ass: richness} must look like Figure~\ref{fig:unawareness_structure}. With more players, which we need in non-trivial matching games, or more characteristics/experiences, they become more complex. However, in our examples we will often just focus on the states that are relevant for the feature we aim to illustrate with the example. 

Let $\mathcal{P}_i$ be the set of all strict preferences over $W \cup \{i\}$ if $i \in M$ and $M \cup \{i\}$ if $i \in W$. For every player $i$ there is a preference mapping defined by $\succ_i: \Omega \longrightarrow \mathcal{P}_i$. That is, every player at each state has preferences over players of the other side and outcomes in which (s)he stays alone.  Since preferences shall only be affected by awareness, we require that each player's preference mapping is constant within each space/awareness level. That is, for any $S \in \mathcal{S}$ and $\omega, \omega' \in S$, ${\succ_i}(\omega) = {\succ_i}(\omega')$. We assume for simplicity that if for $\omega \in S$ and $\omega' \in S'$ we have $j \ {\succ_i}(\omega) \ j'$ and $j \ {\succ_i}(\omega') \ j'$, then $j \ {\succ_i}(\omega'') \ j'$ for $\omega'' \in S \vee S'$. That is, if $i$ prefers $j$ to $j'$ with awareness level $S$ and also with awareness level $S'$, then $i$ also prefers $j$ to $j'$ with the joint awareness level $S \vee S'$. 

Recall that a \emph{matching} is a one-to-one function $\mu: M \cup W \longrightarrow M \cup W$ such that $\mu(m) = w$ if and only if $\mu(w) = m$. If for $i \in M \cup W$, $\mu(i) = i$, then $i$ is unmatched. Let $\mathsf{M}$ denote the set of all matchings. 

Matchings may allow players to discover new characteristics of players or make unanticipated experiences that subsequently change their preferences. That is, matchings may lead to changes awareness and thus beliefs and preferences. This is modeled via a finite state machine $\langle \Omega, \mathsf{M}, \tau \rangle$ with transition function $\tau: \Omega \times \mathsf{M} \longrightarrow \Omega$ defined as follows: 
\begin{itemize}
\item[(i)] For any $\omega \in \check{S}$ and $\mu \in \mathsf{M}$, we require $\tau(\omega, \mu) \in \check{S}$ such that $S_{t_i(\tau(\omega, \mu))} \trianglerighteq S_{t_i(\omega)}$ for  $i \in M \cup W$. That is, every player's awareness can never decrease.
\item[(ii)] We extend $\tau$ to all states in $\Omega$ by for any $S \in \mathcal{S}$, $\omega \in \check{S}$, and $\mu \in \mathsf{M}$, $\tau\left(r^{\check{S}}_S(\omega), \mu\right) = r^{\check{S}}_S\left(\tau(\omega, \mu)\right)$.
\end{itemize}

Note that we allow $i$ not only to become aware from her/his own match but also from a change of the matching that does not involve $i$. Note further that states implicitly encode four features: First, they encode the point-belief of each player via the type-mapping. Point beliefs are correct up to differences in awareness. Second, states encode awareness of each player, also via the type-mapping. The awareness level is given by the space in which the value of the player's type is located. Third, states encode preferences of each player via the preference map. Finally, they encode the transition across states (and thus the change of point-beliefs, awareness, and preferences of each player) conditional on matchings via the finite state machines. Assumption~\ref{ass: redundancies} has implications beyond awareness. For transitions, the assumptions implies that for each profile of awareness and matching, each player has a unique experience and thus a transition to unique potentially different profile of awareness. I.e., the assumption rules out the case where the same profile of awareness and matching can give rise to different experiences for one player. 

\begin{defin} A finite dynamic two-sided matching game with unawareness is defined by $$\langle (\mathcal{S}, \trianglerighteq), (r^{S'}_S)_{S' \trianglerighteq S}, M, W, (t_i)_{i \in M \cup W}, (\succ_i)_{i \in M \cup W}, \mathsf{M}, \tau \rangle.$$ 
\end{defin}

The model introduced so far can be interpreted for instance as modeling unawareness of preference-relevant characteristics of players and their dynamics. E.g., consider a set of preference-relevant characteristics. For each subset of characteristics, there is a state space modeling everything relevant to the players but only pertaining to this subset of characteristics. The lattice order $\trianglerighteq$ on state spaces is induced by set inclusion on the set of characteristics. A state describes now for each player of which characteristics in the subset he/she is aware of. For each space, preferences are constant in states because they are driven by the characteristics associated with the state space. All what differs from state to state is the awareness of characteristics by players. At one state in the state space the player may be aware of all characteristics associated with the space, while at another state of the state space the player may by unaware of some and thus ``live'' in an even less expressive state space. The preference of the player is now given by his/her preference in the less expressive state space.

\section{Self-Confirming Stable Outcomes\label{sec:selfconfirming}} 

In this section, we will define step-by-step our solution concept. Our aim is a solution concept that features a stable matching given beliefs and stable beliefs given the matching. 

For any $m \in M$ and $w, w' \in W$, define $[w \succ_m w']: = \{\omega \in \Omega : w \ {\succ_m}(\omega) \ w'\}$. Analogously, define  $[m \succ_w m']$ for any $w \in W$ and $m, m' \in M$. Given matching $\mu \in \mathsf{M}$, the set of states in which $(m, w)$ forms a \emph{blocking pair} is $[m \succ_w \mu(w)] \cap [w \succ_m \mu(m)]$. Similarly, for any $i \in M \cup W$ and matching $\mu \in \mathsf{M}$, define $[i \succ_i \mu(i)] :=\{ \omega \in \Omega : i \ {\succ_i}(\omega) \ \mu(i)\}$. This is the set of states in which player $i$ prefers to stay alone rather than stay with her/his current partner in the matching $\mu$. 

For any event $E \subseteq \Sigma$ and $i \in M \cup W$, let $K_i(E) = \{\omega \in \Omega : t_i(\omega) \in E\}$, if there exists $\omega \in \Omega$ such that $t_i(\omega) \in E$. Otherwise, let $K_i(E) = \emptyset^{S(E)}$. $K_i(E)$ is the set of states in which player $i$ believes $E$. By the properties of the $t_i$, if $E$ is an event in $\Sigma$, then $K_i(E)$ is an event. For any pair $(m, w)$, let $K_{m, w}(E) = K_m(E) \cap K_w(E)$. That is, $K_{m, w}(E)$ is the event that $E$ is mutual belief among $m$ and $w$. Finally, let $CK_{m, w}(E) = \bigcap_{n \geq 1} K^n_{m, w}(E)$. This is the event that $E$ is common belief among $m$ and $w$. These are the usual (pairwise) mutual belief and common belief operators specialized to our setting. 

In a standard matching model with complete information and full awareness, a matching is stable if there is no blocking pair. In our setting, there might be a blocking pair without common belief thereof. Consequently, they have no agreement to block. We say that in the matching $\mu \in \mathsf{M}$, at state $\omega$ it is common belief among $m$ and $w$ that they form a blocking pair if $\omega \in CK_{m, w}([m \succ_w \mu(w)] \cap [w \succ_m \mu(m)])$.\footnote{We often just refer to ``pairwise common belief'' if the pair of players is clear from the context.} 

\begin{defin}[Stability] We say that $\mu$ is \emph{stable} at $\omega$ if 
\begin{itemize}
\item[(i)] there does not exist $(m, w) \in M \times W$ such that $\omega \in CK_{m, w}([m \succ_w \mu(w)] \cap [w \succ_m \mu(m)])$, and 
\item[(ii)] there does not exist $i \in M \cup W$ such that $\omega \in [i \succ_i \mu(i)]$. 
\end{itemize} 
\end{defin} That is, we say that $\mu$ is stable at $\omega$ if there does not exist a pair $(m, w)$ such that at $\omega$ it is common belief among $m$ and $w$ that $(m, w)$ forms a blocking pair. Moreover, there should also not exist a player who prefers to stay alone over her current match. 

In a standard matching model with complete information and full awareness, stable matchings are in the core of the matching game. In our setting with asymmetric unawareness, stable matchings are in the coarse core. The coarse core has been introduced by Wilson (1978) for exchange economies with asymmetric information. It has been extended to general TU games with incomplete information and unawareness by Bryan et al. (2022). Similar ideas can be used to extend it to NTU games with incomplete information and unawareness like our matching games. 
\begin{figure}[h]
    \centering
    \includegraphics[width=0.6\textwidth]{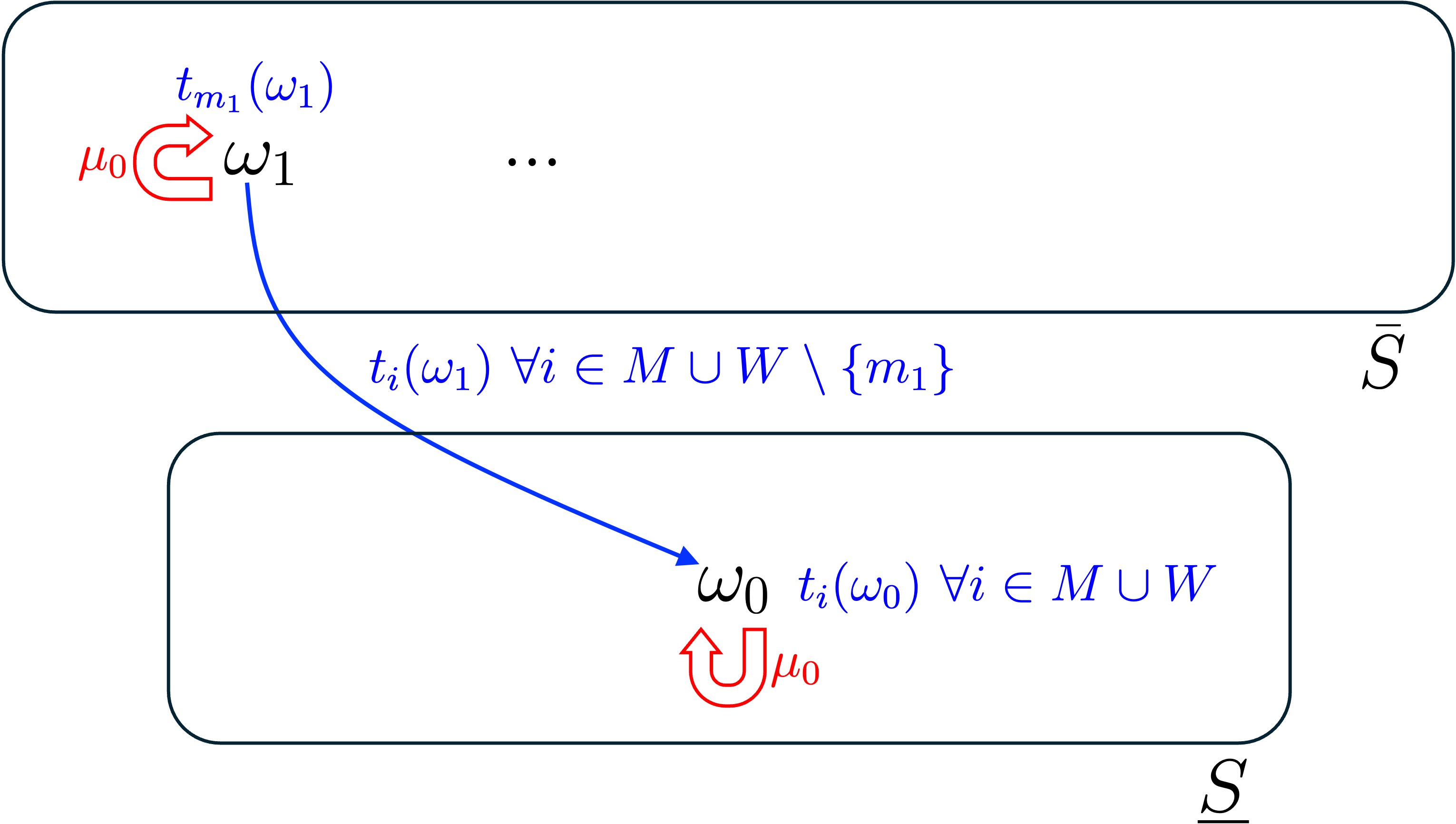}
    \caption{Unawareness Structure of Example 3}
    \label{fig:example_stability_structure}
\end{figure} 

Since our stability notion generalizes absence of blocking pairs to absence of pairwise common belief in blocking, we illustrate this novel feature with a simple example. \\

\noindent \textbf{Example 3 } There are two men and women each, $M = \{m_1, m_2\}$ and $W = \{w_1, w_2\}$. The preference mappings are given by the following rank order lists: 
$$\begin{array}{ccc}
\begin{array}{ll}
\succ_{m_1}: & w_1, w_2\\
\succ_{m_2}: & w_1, w_2\\
\succ_{w_1}: & m_1, m_2\\
\succ_{w_2}: & m_1, m_2\\
\multicolumn{2}{c}{\omega_0}
\end{array} & \quad \quad \quad & 
\begin{array}{ll}
\succ_{m_1}: & w_2, w_1\\
\succ_{m_2}: & w_1, w_2\\
\succ_{w_1}: & m_1, m_2\\
\succ_{w_2}: & m_1, m_2\\
\multicolumn{2}{c}{\omega_1}
\end{array}
\end{array}$$ All players have preferences constant in the states except for man $m_1$. The unawareness structure is depicted in Figure~\ref{fig:example_stability_structure}. There are two spaces, the richer space $\bar{S}$ and the poorer space $\underline{S}$. We just focus on the two states relevant to our example, which is $\omega_1$ in $\bar{S}$ and $\omega_0$ in $\underline{S}$. (We omit the projections as $\omega_1$ projects to $\omega_0$.)  The type mappings for the point beliefs are depicted in blue.\footnote{From now on we suppress circles to avoid clutter.} Importantly, at state $\omega_1$, man $m_1$'s point-belief is at $\omega_1$ while all others have point-belief $\omega_0$. Thus, all except man $m_1$ are unaware. 

\begin{figure}[h]
    \centering
    \includegraphics[width=0.13\textwidth]{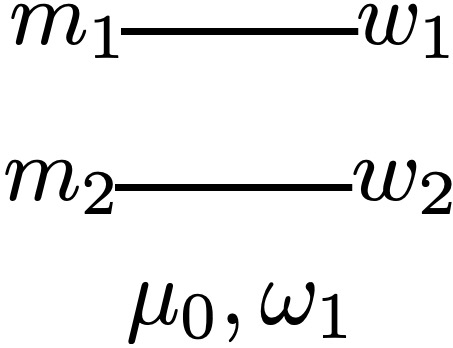}
    \caption{Stable matching at $\omega_1$}
    \label{fig:example_stability_matching}
\end{figure}
Consider now the matching $\mu_0$ given in Figure~\ref{fig:example_stability_matching}. In this matching, man $m_1$ is matched to $w_1$ while at $\omega_1$ he strictly prefers $w_2$ over $w_1$. In fact, they form a blocking pair in the standard sense. Nevertheless, this matching is stable in our sense at $\omega_1$ because at $\omega_1$ it is \emph{not} a common belief among $m_1$ and $w_2$ that they form a blocking pair. This is because woman $w_2$ is unaware and believes that man $m_1$ strictly prefers $w_1$ over herself. This example illustrates the difference between absence of blocking and absence of pairwise common belief in blocking and thus the difference between the standard notion of stability and our notion of stability. \hfill $\Box$\\ 

Stability itself is not a satisfactory solution in our setting because $\mu$ being stable at $\omega$ does not rule out that some players discover something in the matching $\mu$ at $\omega$ that changes their preferences and consequently destabilizes the previously stable matching so that they want to get divorced. For a satisfactory solution, we also need that beliefs are stable given the matching.

\begin{defin} We say that a state $\omega \in \Omega$ is \emph{absorbing} given $\mu$ if $\tau(\omega, \mu) = \omega$. 
\end{defin}

Putting these two ideas together yields our solution concept: 

\begin{defin}[Self-confirming outcome] We say that an outcome $(\omega, \mu)$ is \emph{self-confirming} if 
\begin{itemize} 
\item[(i)] $\mu$ is stable at $\omega$, and 
\item[(ii)] $\omega$ is absorbing given $\mu$. 
\end{itemize} 
\end{defin} 
\begin{figure}[h]
    \centering
    \includegraphics[width=0.6\textwidth]{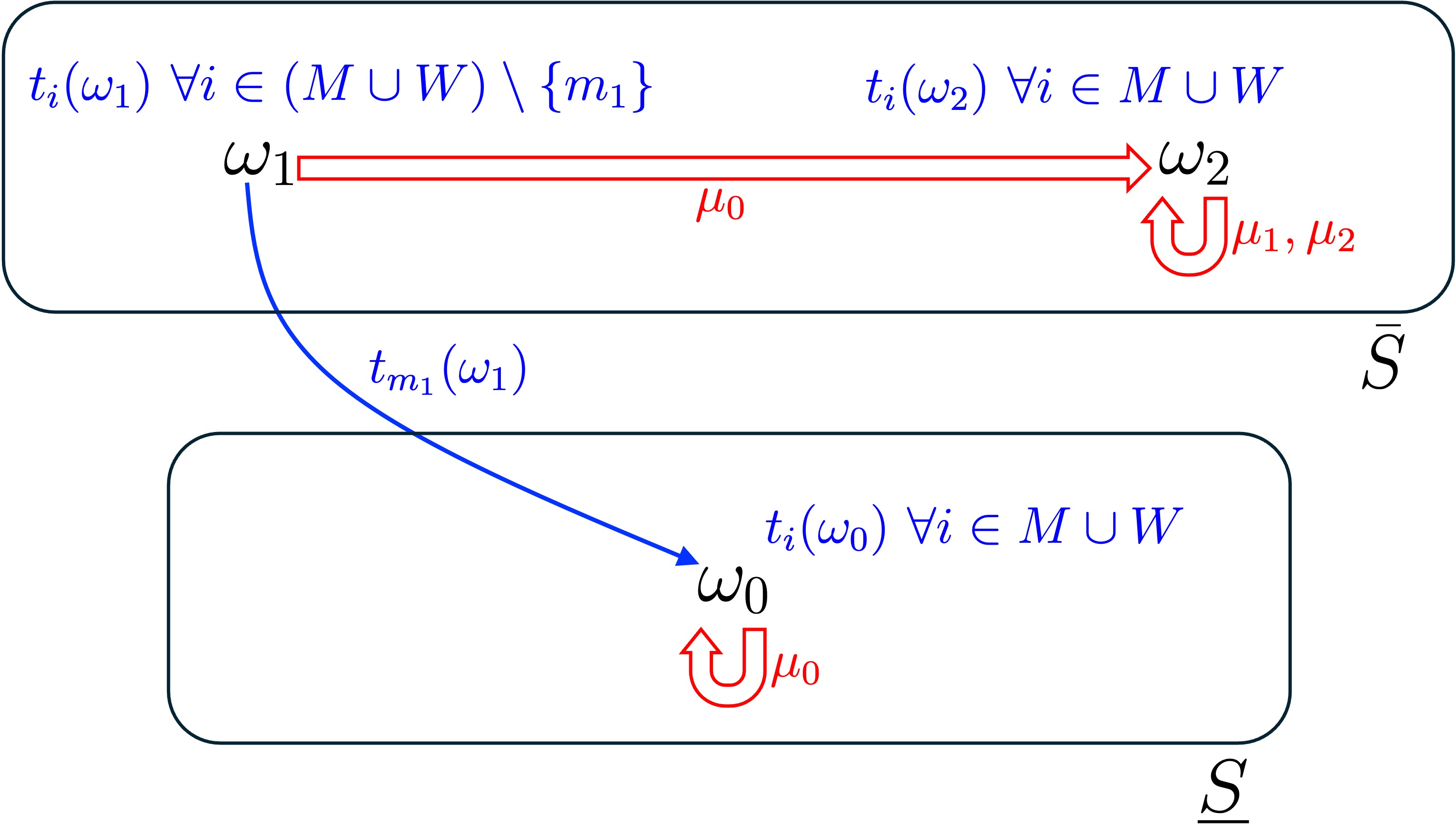}
    \caption{Unawareness Structure of Example 4}
    \label{fig:example_absorbing}
\end{figure} 

Since the absorbing state given the matching is a novel feature of the solution concept, we illustrate it with the following simple example.\\ 

\noindent \textbf{Example 4 } There are two men and women each, $M = \{m_1, m_2\}$ and $W = \{w_1, w_2\}$. The preference mappings are given by the following rank order lists: 
$$\begin{array}{ccc}
\begin{array}{ll}
\succ_{m_1}: & w_1, w_2\\
\succ_{m_2}: & w_1, w_2\\
\succ_{w_1}: & m_1, m_2\\
\succ_{w_2}: & m_1, m_2\\
\multicolumn{2}{c}{\omega_0}
\end{array} & \quad \quad \quad & 
\begin{array}{ll}
\succ_{m_1}: & w_2, w_1\\
\succ_{m_2}: & w_1, w_2\\
\succ_{w_1}: & m_1, m_2\\
\succ_{w_2}: & m_1, m_2\\
\multicolumn{2}{c}{\omega_1, \omega_2}
\end{array}
\end{array}$$ All players have preferences constant in the states except for man $m_1$. The unawareness structure is depicted in Figure~\ref{fig:example_absorbing}. There are two spaces, the richer space $\bar{S}$ and the poorer space $\underline{S}$. We focus on two states in the richer spaces, $\omega_1$ and $\omega_2$, and one state in the poor space, $\omega_0$. (We omit the projections as they are trivial. Both states $\omega_1$ and $\omega_2$ project to $\omega_0$.) The type mappings for the point beliefs are depicted in blue. Importantly, at state $\omega_1$, man $m_1$'s point-belief is at $\omega_0$. That is, he is unaware of something that obviously effects his preferences because his preference in states in $\bar{S}$ differs from his preference at $\underline{S}$. 
\begin{figure}[h]
    \centering
    \includegraphics[width=0.8\textwidth]{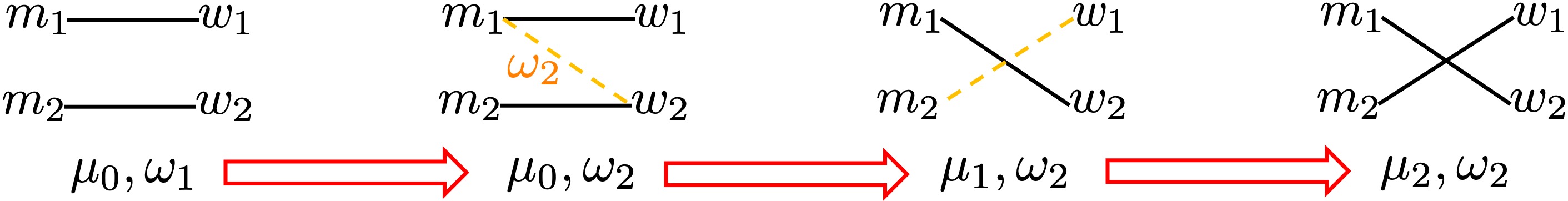}
    \caption{Process of Example 4}
    \label{fig:example_absorbing_sequence}
\end{figure}

The relevant state transitions are depicted by red arrows in Figure~\ref{fig:example_absorbing}. We print the matching(s) that facilitate the state transitions beside the red arrows. To make the process transparent, we depict the process separately by Figure~\ref{fig:example_absorbing_sequence}. Let $\omega_1$ be the initial state. At that state all players except $m_1$ believe in $\omega_1$. Man $m_1$ believes in state $\omega_0$. Let matching $\mu_0$ be the initial matching. It is easy to verify that it is stable given $\omega_1$ and also stable given $\omega_0$ as there is no pair who has common belief in blocking. Yet, as argued above, stability is not sufficient for a satisfactory solution concept. Given state $\omega_1$ and matching $\mu_0$, man $m_1$ becomes aware as the process transitions to state $\omega_2$ at which $m_1$'s point-belief is correctly $\omega_2$. (Note that initially at $\omega_0$ man $m_1$ did not anticipate this transition.) Consequently, $m_1$'s preference changes. At $\omega_2$, $m_1$ and $w_2$ now form a blocking pair, this is common belief (not just among $m_1$ and $w_2$ but among all players in this case), and thus $\mu_0$ is not stable anymore. When satisfying the blocking pair, both $m_1$ and $w_2$ have to get divorced from their current partners and we reach matching $\mu_1$ (at state $\omega_2$). This matching is not stable since both $m_2$ and $w_1$ prefer to be matched to each other rather than staying single. Thus they form a blocking pair and this is common belief. Satisfying this blocking pair, we reach matching $\mu_2$ (at $\omega_2$). This matching is stable given $\omega_2$ as their no blocking pairs. Moreover, state $\omega_2$ is absorbing given $\mu_2$, i.e., awareness or preferences do not change given $\mu_2$. This example shows that stability of matching at $\omega_1$ is not enough for a solution concept. Our solution concept also requires the state to be absorbing, like at outcome $(\mu_2, \omega_2$). \hfill $\Box$

\begin{prop}\label{prop: existence}  Every finite dynamic two-sided matching game with unawareness has a self-confirming outcome. 
\end{prop}

\noindent \textsc{Proof. } (i): Call the two-sided matching game with unawareness at a \emph{given} state the stage game. Any stage game of the two-sided matching game with unawareness involves the same finite number of finite state spaces. Any finite state machine on a finite space must have absorbing sets. What is left to show is that there is an absorbing set that is a singleton: Consider $\omega \in \check{S}$ such that $S_{t_i(\omega)} = \check{S}$ for all $i \in M \cup W$. Such a state exists in $\check{S}$ by Assumption~\ref{ass: richness}. We have $\tau_i(\omega, \mu) = \omega$ for all $\mu$. To see this, suppose to the contrary that $\tau_i(\omega, \mu) = \omega'$ for some $\omega' \neq \omega$ and $\mu \in \mathsf{M}$. By Assumption~\ref{ass: redundancies}, there exists $i \in M \cup W$ such that $S_{t_i(\omega')} \neq S_{t_i(\omega)}$. Since $S_{t_i(\omega)} = \check{S}$, we must have $S_{t_i(\omega')} \triangleleft \check{S}$. But this contradicts that assumption that $S_{t_i(\tau(\omega, \mu))} \trianglerighteq S_{t_i(\omega)}$. We conclude that $\omega$ is absorbing. 

(ii): Every state $\omega \in \Omega$ pins down a strict preference profile of all agents. Consider the absorbing state from part (i).  By Gale and Shapley (1962), a stable matching in the standard sense exists, i.e. a matching without a blocking pair. Consequently, there can also be no common belief in blocking. Hence, it is stable.

By (i) we argued that there exists an absorbing state $\omega$. By (ii), there is a stable match at $\omega$. Hence, there exists a self-confirming outcome.\hfill $\Box$\\

The proof is straightforward: Awareness can only go up. Once all players are aware of everything, there must exist an absorbing state.  At that state there must exist a stable match in the standard sense of Gale and Shapley (1962). Absence of blocking pairs means also no common belief in blocking. Hence, it is also stable in our sense. 
\begin{figure}[h]
    \centering
    \includegraphics[width=0.6\textwidth]{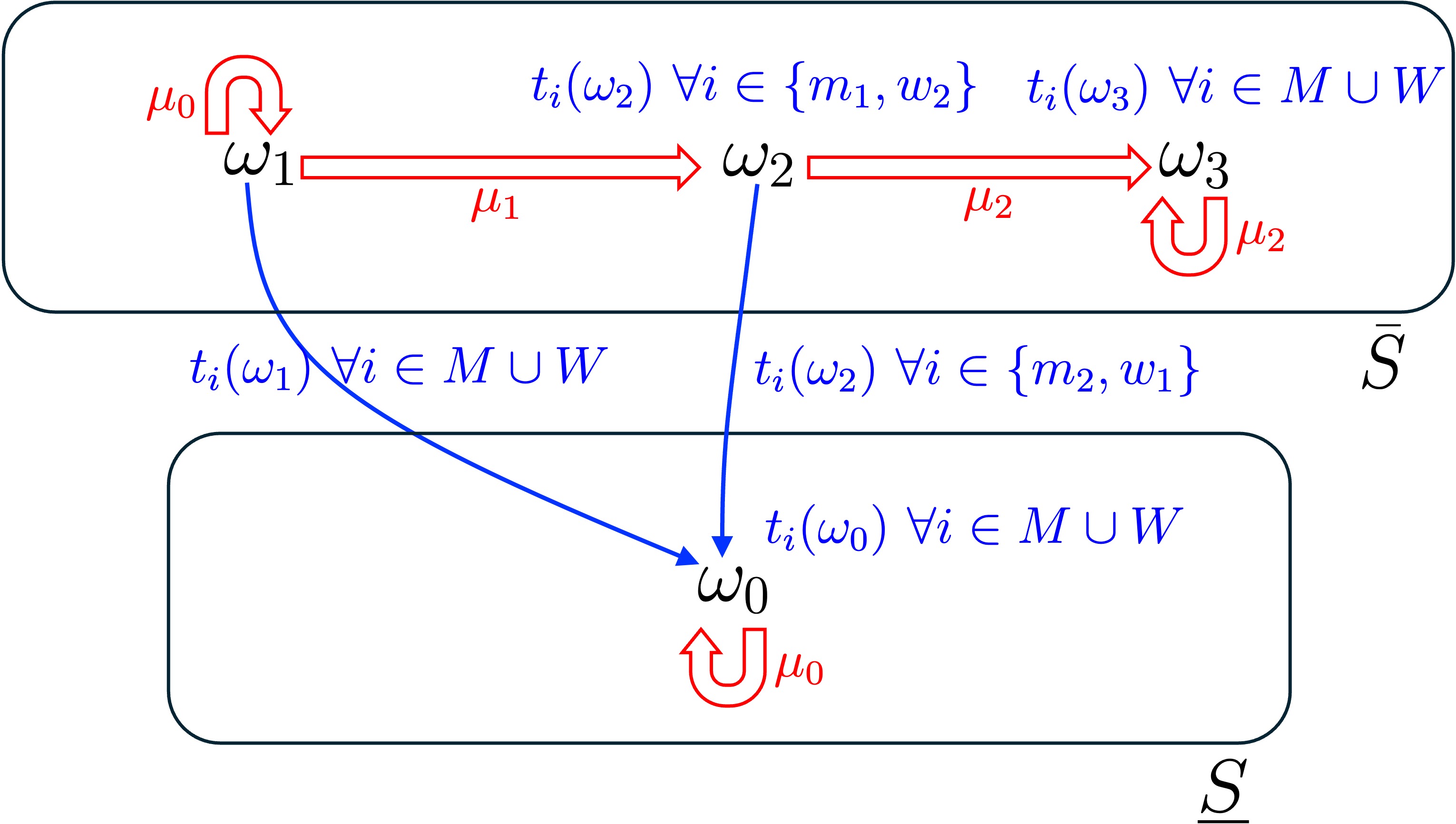}
    \caption{Trapped in Unawareness by Marriage}
    \label{fig:example_trap}
\end{figure}

While for showing existence it is enough to argue with the upmost space, there can be absorbing states and self-confirming outcomes that involve unawareness. In the following simple example we illustrate how players are trapped by marriage in their unawareness.\\

\noindent \textbf{Example 5 }  There are two men and women each, $M = \{m_1, m_2\}$ and $W = \{w_1, w_2\}$. The preference mappings are given by the following rank order lists: 
$$\begin{array}{ccc}
\begin{array}{ll}
\succ_{m_1}: & w_1, w_2\\
\succ_{m_2}: & w_1, w_2\\
\succ_{w_1}: & m_1, m_2\\
\succ_{w_2}: & m_1, m_2\\
\multicolumn{2}{c}{\omega_0}
\end{array} & \quad \quad \quad & 
\begin{array}{ll}
\succ_{m_1}: & w_2, w_1\\
\succ_{m_2}: & w_1, w_2\\
\succ_{w_1}: & m_2, m_1\\
\succ_{w_2}: & m_1, m_2\\
\multicolumn{2}{c}{\omega_1, \omega_2, \omega_3}
\end{array}
\end{array}$$
The unawareness structure is given by Figure~\ref{fig:example_trap}. At $\omega_1$, all players' point-belief is $\omega_0$ in $\underline{S}$. That is, all players are unaware. Let $\mu_0$ as in the example earlier, i.e., the matching given in Figure~\ref{fig:example_stability_matching}. This matching is stable given $\omega_1$ because at that state all players are unaware and their point-belief is $\omega_0$. Thus, there is no pair who has common belief in blocking. Since the $\omega_1$ is absorbing given $\mu_0$, the outcome $(\mu_0, \omega_1)$ is a self-confirming outcome despite all players being unaware. We also observe in Figure~\ref{fig:example_trap} that there would be other matchings like $\mu_1$ that would allow $m_1$ and $w_2$ to become aware (and subsequently matching $\mu_2$ that allow the remaining players to become aware). However, since $(\mu_0, \omega_1)$ is absorbing, these states are not reached. Players are trapped in unawareness in the marriages. We will return to this example later in Subsection~\ref{sec:infidelity} when discussing infidelity. \hfill $\Box$\\

More interesting than showing existence of self-confirming outcomes is the process of finding those self-confirming outcomes with transformative experiences and discoveries in matchings, leading to changes of preferences, further attempts of trying to find stable matches, further preference-perturbing discoveries etc. We are interested in a natural decentralized matching process that leads to self-confirming outcomes. From Section~\ref{sec: revisited} we know that even without unawareness, it is not straightforward for decentralized matching processes to reach a stable matching. With asymmetric unawareness, changes of awareness may also lead to changes of preferences which complicates the process even further. 

For any $\omega \in \Omega$, $\mu \in \mathsf{M}$ and $w \in W$, let 
$$M_w(\omega, \mu) := \left\{m \in M : \omega \in CK_{m, w}([m \succ_w \mu(w)] \cap [w \succ_m \mu(m)])\right\}  \cup \{w: w \ {\succ_w}(t_w(\omega)) \ \mu(w)\}.$$ 
The first term, $\left\{m \in M : \omega \in CK_{m, w}([m \succ_w \mu(w)] \cap [w \succ_m \mu(m)])\right\}$, is the set of men $m$ for which it is common belief among $w$ and $m$ at $\omega$ and matching $\mu$ that $m$ and $w$ form a blocking pair. The second term, $\{w: w \ {\succ_w}(t_w(\omega)) \ \mu(w)\}$ is nonempty only if woman $w$ prefers to stay alone at $\omega$ rather than with her partner in the matching $\mu$. 

We say that at $\omega$ and $\mu$, $w$ and $i$ are a $(w, i)$-commonly believed $w$-best blocking pair if $i \in M_w(\omega, \mu)$ and $i \ {\succ_w}(t_w(\omega)) \ j$ for any $j \in M_w(\omega, \mu)$ with $j \neq i$. Analogously, define $W_m(\omega, \mu)$ and commonly believed $m$-\emph{best blocking} pairs. We say that at $\omega$ and $\mu$, $m$ and $w$ are a $(i, j)$-commonly believed \emph{mutual best blocking} pair if it is both a $(i, j)$-commonly believed $i$-best blocking pair and $(i, j)$-commonly believed $j$-best blocking pair. Of course, our earlier assumption ensures that either $i$ and $j$ are members of different sides of the market or $i = j$. Denote by $B(\omega, \mu) \subseteq M \times W$ the commonly believed best blocking pairs at $\omega$ and $\mu$. Denote by $MB(\omega, \mu) \subseteq M \times W$ the commonly believed mutual best blocking pairs at $\omega$ and $\mu$. 

\begin{lem} For any $\omega$ and $\mu$, if $\mu$ is not stable at $\omega$, then $B(\omega, \mu) \neq \emptyset$. 
\end{lem}

\noindent \textsc{Proof.} For any $\omega \in \Omega$, if $\mu$ is not stable, then there exists $i \in W $ such that $M_i(\omega, \mu) \neq \emptyset$ or $j \in M$ such that $W_j(\omega, \mu) \neq \emptyset$ . Note also that since $M_i(\omega, \mu)$ and $W_j(\omega, \mu) $ are finite, the commonly believed $i$-best blocking pair or the commonly believed $j$-best blocking pair exist. Therefore, $B(\omega, \mu) \neq \emptyset$. \hfill $\Box$\\

In contrast to $B(\omega, \mu)$, the set $MB(\omega, \mu)$ may be empty even if $\mu$ is not stable. Therefore, define 
\begin{eqnarray*} \widehat{MB}(\omega, \mu) & := & \left\{ \begin{array}{ll} MB(\omega, \mu) & \mbox{ if } MB(\omega, \mu) \neq \emptyset \\
B(\omega, \mu) & \mbox{ otherwise.} \end{array} \right. 
\end{eqnarray*} 

Fix small $\varepsilon \in (0, 1)$ and define a matching process by transition probabilities $P^{\varepsilon}$ such that for any $\omega$ and $\mu$, 
\begin{eqnarray*} P^{\varepsilon}(\mu', \omega' \mid \mu, \omega) & := & \left\{ \begin{array}{cl} \frac{1 - \varepsilon}{|\widehat{MB}(\omega, \mu)|}  & \mbox{if } \mu' \mbox{ differs from } \mu \mbox{ by satisfaction of exactly one} \\ & \mbox{pair in } \widehat{MB}(\omega, \mu) \mbox{ and } \tau(\omega, \mu') = \omega'; \\ \\ 
\frac{\varepsilon}{|B(\omega, \mu)|} & \mbox{if } \mu' \mbox{ differs from } \mu \mbox{ by satisfaction of exactly one} \\ & \mbox{pair in } B(\omega, \mu) \mbox{ and } \tau(\omega, \mu') = \omega'; \\ \\ 
1 &  \mbox{if }  \mu' = \mu \mbox{ and } \mu \mbox{ is stable given } \omega \mbox{ and } \tau(\omega, \mu') = \omega'; \\ \\
0 & \mbox{else. } \end{array} \right.  
\end{eqnarray*} 

We define the matching process by transition probabilities on the product space $\Omega \times \mathsf{M}$. This warrants some explanations. Naturally, the process would proceed as follows: Given a current state $\omega$ and awareness/preferences at $\omega$, the current match $\mu$ may feature some blocking pairs w.r.t. preferences at $\omega$. Satisfying some blocking pair (i.e., preferably some mutual optimal blocking pair) would lead to another match $\mu'$. At this match $\mu'$ and state $\omega$, there may be discoveries leading to another state $\tau(\omega, \mu') = \omega'$. At this state and corresponding awareness/preferences, there might be blocking pairs. Satisfying a blocking pair may lead to yet another matching $\mu''$ ... While in this explanation, we let states transit \emph{after} the transition of the matches, the above formal description of the process involves a simultaneous transition of states and matches. It can be thought of just describing the process at every ``even'' period. This is well-defined because states move deterministically according to a finite state machine given the prior state and the new match. The process on $\Omega \times \mathsf{M}$ is not deterministic though because blocking pairs are selected randomly with $(1 - \varepsilon)$ assigned to mutual best blocking pairs if they exist. 

\begin{prop}\label{prop: convergence}  For any (initial) outcome $(\mu, \omega)$ and $\varepsilon \in (0, 1)$, the matching process $P^{\varepsilon}$ convergences almost surely to a self-confirming outcome. 
\end{prop}

\noindent \textsc{Proof. } First, we argue that each absorbing set of $P^{\varepsilon}$ cannot involve more than one state. To see this, consider an absorbing set of $P^{\varepsilon}$ that involves at least two states. Denote them by $\omega^*$ and $\omega^{**}$. By the definition of absorbing set, each outcome (consisting both of a state and a matching) is reachable from any other outcomes in the absorbing set via a finite number of transitions. Thus, $\omega^*$ must be reachable from $\omega^{**}$ and vice versa. Since for each agent $i$, awareness can only increase along the process, we must have $S_{t_i(\omega^*)} = S_{t_i(\omega^{**})}$ for all $i$. The assumption of no redundancies, Assumption~\ref{ass: redundancies}, implies now that $\omega^* = \omega^{**}$. 

Second, we argue that each absorbing set of $P^{\varepsilon}$ cannot involve more than one matching and that matching must be stable. Suppose to the contrary that there exists an absorbing set that involves an unstable matching denoted by $\mu_1$.\\ 

\noindent \textbf{Claim:} Consider w.l.o.g. $\mu_1$ in the absorbing set for $\omega^*$. We claim that there exists a sequence of matchings $\mu_1, ..., \mu_k$, with $k \leq 2|W| \cdot |M|$, such that $\mu_k$ is stable, and for each $i = 1, ..., k - 1$, there exists a pairwise commonly believed best blocking pair $(m_i, w_i)$ such that $\mu_{i + 1}$ is obtained from $\mu_i$ by satisfying $(m_i, w_i)$.

To prove the claim note first that the preference profile is fixed by the absorbing state $\omega^*$. For any women $w_i$, since the preference of women $w_i$ is an ordered list over $M \cup \{w_i\}$ of length $|M| + 1$, we consider the order from the worst to the best, with the worst ranking $0$ and the best has rank $|M|$. Define the payoff of agent $w_i$ in matching $\mu$ by $p_{w_i}(\mu) = k \in \{0, ... , |M|\}$ if $\mu(w_i)$ is at the $k$th place of the order. That is, if $w_i$ is matched to her most preferred agent in $M \cup \{w_i\}$, then here payoff is $|M|$. 

Let $X(\mu)$ be the set of matched women at $\mu$. Define a ``potential'' function $\Phi(\mu) = \sum_{w \in X(\mu)}(|M| - p_w(\mu))$. Note that this potential function is minimized when all women are matched to their ``best'' agent on their lists, respectively. Function $\Phi$ is bounded above by $|W| \cdot |M|$ as $|X(\mu)| \leq |W|$ and $p_w(\mu) \geq 0$ for $w \in X(\mu)$.

To construct a sequence of matchings $\mu_1, ..., \mu_k$, we divide it into two phases. The first phase features the sequence $\mu_1, ..., \mu'$ and the second $\mu', ..., \mu_k$. In the first phase, we move from $\mu_i$ to $\mu_{i+1}$ by satisfying pairwise commonly believed best blocking pairs that involve matched women only. With each satisfied pairwise commonly believed best blocking pair, $\Phi$ decreases by at least 1 because: (1) Starting from $\mu_i$, if this woman gets rematched to an unmatched man, the set $X(\mu_{i+1}) = X(\mu_i)$, and her payoff increases. (2) If she gets rematched to a matched man, then $X(\mu_{i+1}) \subsetneqq X(\mu_i)$ (i.e., her new matching partner leaves his current partner), and her payoff increases. (3) If she becomes unmatched, her term in the potential function is dropped, and that term must have been positive before. Hence, at most after $|W| \cdot |M|$ steps, no matched woman can improve her payoff, which implies that no matched woman has a commonly believed blocking pair. Furthermore, if there is no pairwise commonly believed blocking pair, there is no pairwise commonly believed best blocking pair. The first phase terminates with a matching $\mu'$ in which no matched woman has a pairwise commonly believed (best) blocking pair. Observe that the process $P^{\varepsilon}$ allows for above sequence of pairwise commonly believed best blocking pairs to be satisfied because at any step the process puts strict positive probability on any pairwise commonly believed best blocking pair at that step. Moreover, we assumed that $(\mu_1, \omega^*)$ is in the absorbing set of $P^{\varepsilon}$. Since we reached $(\mu', \omega^*)$ with $P^{\varepsilon}$, it implies that $(\mu', \omega^*)$ is in the absorbing set of $P^{\varepsilon}$ as well. 

In the second phase, suppose we start from the matching $\mu'$. If there is no pairwise commonly believed blocking pair in $\mu'$ (also among unmatched women), then let $\mu'=\mu_k$ and the second phase terminates. Otherwise, the second phase continues as follows. Since no matched woman has a pairwise commonly believed blocking pair in $\mu'$, $\{(m, w): \mu'(w) \neq w, \omega^* \in CK_{m,w}([m \succ_w \mu'(w)] \cap [w \succ_m \mu'(m)])\} = \emptyset$. That is, the set of pairwise commonly believed blocking pairs that involves a matched woman is empty. 

Satisfy a pairwise commonly believed best blocking pair of an unmatched woman, $(m^*,w^*)$, and call the resulting matching $\mu''$. We argue that in $\mu''$, no matched woman can have a pairwise commonly believed blocking pair, i.e. $\{(m, w): \mu''(w) \neq w, \omega^* \in CK_{m,w}([m \succ_w \mu''(w)] \cap [w \succ_m \mu''(m)])\} = \emptyset$. Since the set of matched woman only change by adding $w^*$ and removing $\mu'(m^*)$ if $m^*$ is matched under $\mu'$, the set of matched women is now $\{w:\mu''(w)\neq w\} =  \{w^*\} \cup \{w: \mu'(w)\neq w\} \setminus \{\mu'(m^*): \mu'(m^*)\neq m^*\}$. Therefore, the set of pairwise commonly believed blocking pairs that involves a matched woman is now
\begin{eqnarray*} \lefteqn{\{(m, w): \mu''(w) \neq w, \omega^* \in CK_{m,w}([m \succ_w \mu''(w)] \cap [w \succ_m \mu''(m)])\} = } \\ & &  
\{(m, w): w = w^*, \omega^* \in CK_{m,w}([m \succ_w \mu''(w)] \cap [w \succ_m \mu''(m)])\} \cup \\ & &
\{(m, w): \mu'(w) \neq w, \omega^* \in CK_{m,w}([m \succ_w \mu''(w)] \cap [w \succ_m \mu''(m)])\} \setminus \\ & &  \{(m, w): w=\mu'(m^*)\neq m^*, \omega^* \in CK_{m,w}([m \succ_w \mu''(w)] \cap [w \succ_m \mu''(m)])\}
\end{eqnarray*} 
We show that the r.h.s. is empty. Consider first the set $$\{(m, w): w = w^*, \omega^* \in CK_{m,w}([m \succ_w \mu''(w)] \cap [w \succ_m \mu''(m)])\}.$$ For $w^*$, since $(m^*, w^*)$ is her pairwise commonly believed best blocking pair given $\mu'$, she cannot have any pairwise commonly believed blocking pair after satisfying $(m^*,w^*)$, i.e., $\{(m, w): w = w^*, \omega^* \in CK_{m,w}([m \succ_w \mu''(w)]\cap[w\succ_m \mu''(m)])\} = \emptyset$.

Next consider the set \begin{align*} & \{(m, w): \mu'(w) \neq w, \omega^* \in CK_{m,w}([m \succ_w \mu''(w)] \cap [w \succ_m \mu''(m)])\} \setminus \\   & \{(m, w): w=\mu'(m^*)\neq m^*, \omega^* \in CK_{m,w}([m \succ_w \mu''(w)] \cap [w \succ_m \mu''(m)])\}.\end{align*} This set concerns pairs that involve matched women under $\mu'$ except the women that was possibly matched under $\mu'$ to the man that got rematched under $\mu''$.  

Recall that for all $m \neq m^*$ we have $\mu''(m)= \mu'(m)$. Moreover, $\mu''(m) \ {\succ_{m^*}}(t_{m^*}(\omega^*)) \ \mu'(m)$. Thus, every man believes that he gets a weakly better match under $\mu''$. Thus, we have that for all $m \in M$, $\{w: w \ {\succ_m}(t_m(\omega^*)) \ \mu''(m)\} \subseteq \{w: w \ {\succ_m}(t_{w}(\omega^*)) \ \mu'(m)\}$ (with strict ``$\subset$'' for $m^*$ and ``$=$'' for all $m \neq m^*$). Similarly, for all $w$ such that $w \neq w^*$ and $w \neq \mu'(m^*)$ (if $\mu'(m^*)\neq m^*$), we have $\mu''(w)= \mu'(w)$. Thus, every woman who is matched in $\mu''$ believes that she gets a weakly better match under $\mu''$ than under $\mu'$. Therefore, for all $w$ with $\mu'(w)\neq w$ and $w \neq \mu'(m^*)$, $\{m: m \ {\succ_w}(t_w(\omega^*)) \ \mu''(w)\} \subseteq \{m: m \ {\succ_w}(t_w(\omega^*)) \ \mu'(w)\}$. 

Therefore, if there exists a pair $(m, w)$ with $w \neq \mu'(w)$ and $w \neq \mu'(m^*)$ such that $\omega^* \in CK_{m,w}([m \succ_w \mu''(w)] \cap [w \succ_m \mu''(m)])\}$, then by the arguments above we must have $\omega^* \in CK_{m,w}([m \succ_w \mu'(w)] \cap [w \succ_m \mu'(m)])\}$. However, this contradicts our earlier conclusion from the first phase that $\{(m, w): \mu'(w) \neq w, \omega^* \in CK_{m,w}([m \succ_w \mu'(w)] \cap [w \succ_m \mu'(m)])\} = \emptyset$. We conclude that $\{(m, w): \mu''(w) \neq w, \omega^* \in CK_{m,w}([m \succ_w \mu''(w)] \cap [w \succ_m \mu''(m)])\} = \emptyset.$

The analogous arguments apply inductively to next steps of the second phase. Since only unmatched woman are able to block in the second phase, men are never left and can only improve, which can only happen at most $|W|\cdot|M|$ times. When the second phase terminates, there is no pairwise commonly believed blocking pair, and the matching is stable given $\omega^*$. Observe that the process $P^{\varepsilon}$ allows for a second-phase sequence of pairwise commonly believed best blocking pairs to be satisfied because at any step the process puts non-zero probability on any pairwise commonly believed best blocking pair at that step. Moreover, we have already shown that $(\mu', \omega^*)$ is in the absorbing set of $P^{\varepsilon}$. Since we reached $(\mu_k, \omega^*)$ with $P^{\varepsilon}$, it implies that $(\mu_k, \omega^*)$ is in the absorbing set of $P^{\varepsilon}$ as well. This completes the proof of the claim.\\  

The claim implies that each absorbing set of $P^{\varepsilon}$ must be a singleton $(\mu, \omega)$ consisting of an absorbing state $\omega$ given $\mu$ and a stable $\mu$ given $\omega$. To see this, note that each absorbing set cannot have more than one matching. Otherwise, each such matchings must be reachable by the process from each other with a finite number of transitions. This would imply that all these matchings of the absorbing set are unstable. However, the claim shows that there is a sequence of matchings allowed by the process leading to a stable matching, a contradiction.\\ 

\noindent \textbf{Claim:} An outcome $(\mu, \omega)$ is self-confirming if and only if it is an absorbing outcome of $P^{\varepsilon}$. 

To prove the claim, consider first ``$\Leftarrow$'' direction: Let $(\mu, \omega)$ be an absorbing outcome of $P^{\varepsilon}$. Then $P^{\varepsilon}(\mu, \omega \mid \mu, \omega) =1 $. By the construction of $P^{\varepsilon}$,  $\mu$ is stable given $\omega$ and $\omega$ is absorbing given $\mu$. Thus, $(\mu, \omega)$ is self-confirming.

``$\Rightarrow$'': Let $(\mu, \omega)$ be self-confirming. Then $\mu$ is stable at $\omega$, and  $\omega$ is absorbing given $\mu$.  Hence, $P^{\varepsilon}(\mu, \omega \mid \mu, \omega) = 1 $, i.e.  $(\mu, \omega)$ be an absorbing outcome of $P^{\varepsilon}$. This completes the proof of the claim.\\ 

To complete the proof of Proposition~\ref{prop: convergence}, it is enough to note that $P^{\varepsilon}$ must converge almost surely to an absorbing outcome. \hfill $\Box$\\

The proof shows first that any absorbing set must exactly involve a single state because awareness must be constant within each absorbing set. Second, it shows that any absorbing set can at most involve one matching and this matching must be stable given the state. This part of the proof slightly extends an argument by Ackermann et al. (2008) to a sequence of satisfying \emph{pairwise commonly believed} best blocking pairs.

\section{Allowing for Flirting\label{sec:flirting}} 

The prior solution concept features very conservative blocking behavior. Only if there is common belief among a pair that they want to block, there is a chance that the pair is selected for blocking. There are situations in which for instance man $m$ believes that although there is currently absence of common belief in blocking with women $w$, if he were to talk to her and raise her awareness, it would result in common belief in blocking. That is, communication involving raising awareness is natural in this setting. Such a communication could be a feature of flirting behavior. Flirting can destabilize outcomes in two ways: First, raising awareness may change preferences and thus allow for blocking pairs. This is illustrated with the help of the following example. \\
\begin{figure}[h]
    \centering
    \includegraphics[width=0.6\textwidth]{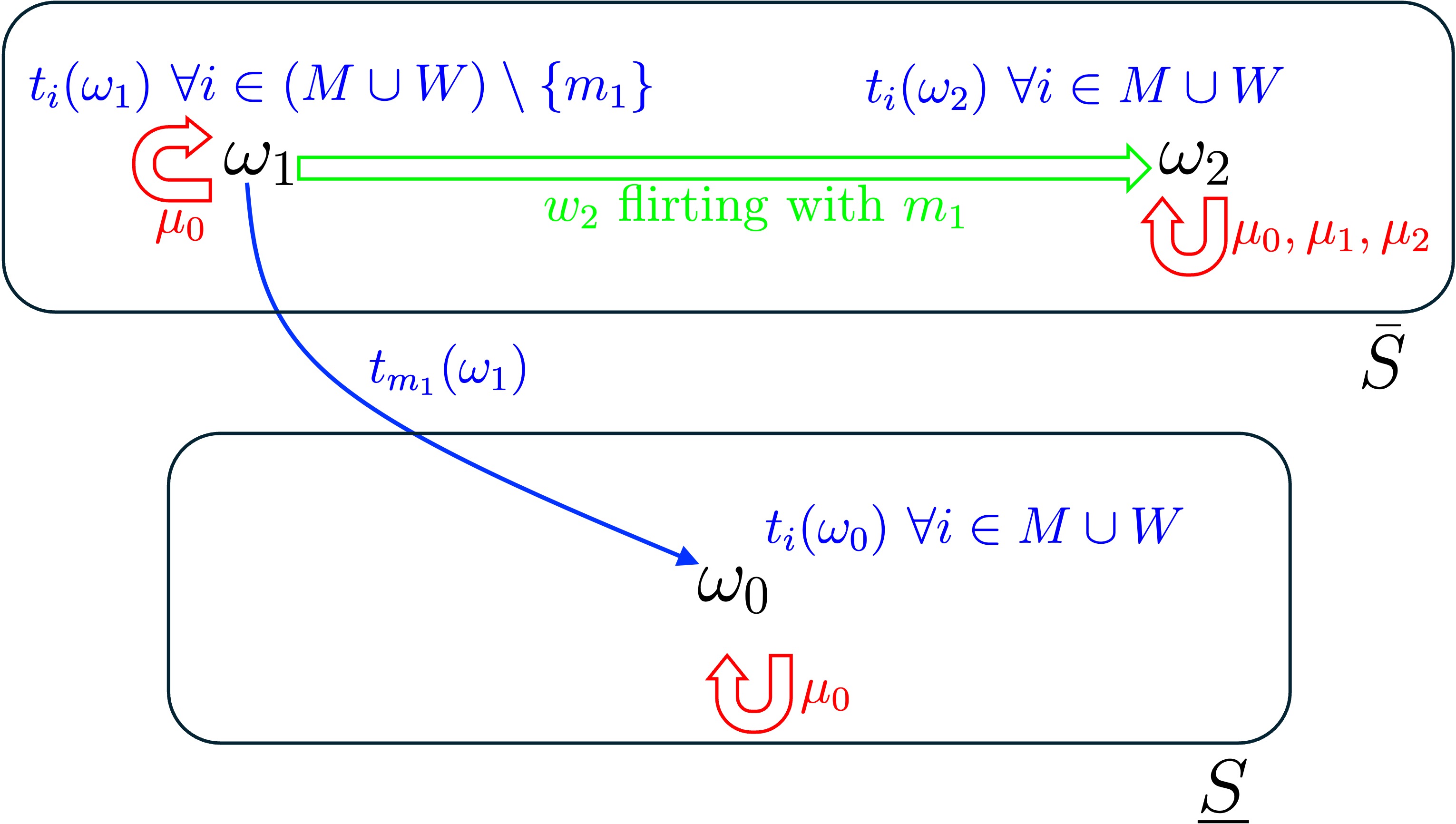}
    \caption{Unawareness Structure of Example 6}
    \label{fig:example_flirting_preference}
\end{figure}

\begin{figure}[h]
    \centering
    \includegraphics[width=0.8\textwidth]{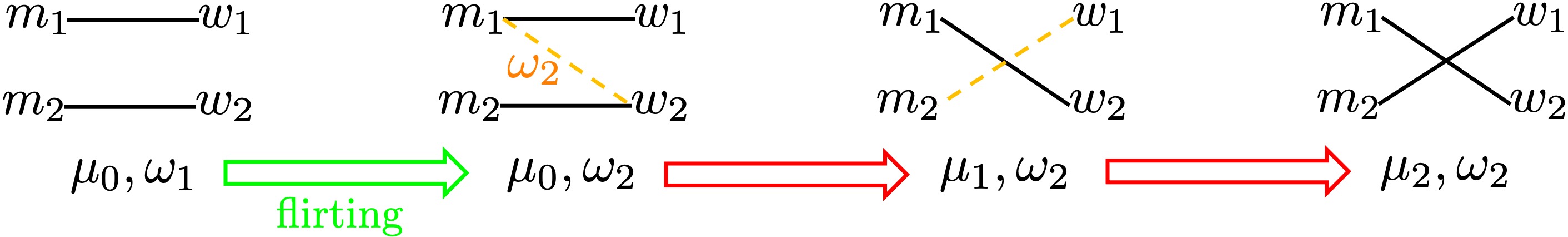}
    \caption{Process of Example 6}
    \label{fig:example_flirting_preference_process}
\end{figure}

\noindent \textbf{Example 6 } This is a variation of prior Example 4. There are two men and women each, $M = \{m_1, m_2\}$ and $W = \{w_1, w_2\}$. The preference mappings are as before given by the following rank order lists: 
$$\begin{array}{ccc}
\begin{array}{ll}
\succ_{m_1}: & w_1, w_2\\
\succ_{m_2}: & w_1, w_2\\
\succ_{w_1}: & m_1, m_2\\
\succ_{w_2}: & m_1, m_2\\
\multicolumn{2}{c}{\omega_0}
\end{array} & \quad \quad \quad & 
\begin{array}{ll}
\succ_{m_1}: & w_2, w_1\\
\succ_{m_2}: & w_1, w_2\\
\succ_{w_1}: & m_1, m_2\\
\succ_{w_2}: & m_1, m_2\\
\multicolumn{2}{c}{\omega_1, \omega_2}
\end{array}
\end{array}$$ The unawareness structure is depicted in Figure~\ref{fig:example_flirting_preference}. The type mappings are as in Example 4. What differs are the state transitions. Recall that $\mu_0$ is stable given $\omega_1$ in Example 4. Now we have also that $\omega_1$ is absorbing given $\mu_0$ as shown by the red arrow from $\omega_1$ to itself. Thus, $(\mu_0, \omega_1)$ is a self-confirming outcome. However, it can be destabilized by flirting. Observe that woman $w_2$ has an incentive to raise man $m_1$'s awareness and thus changing his preference in her favor. This is indicated in Figure~\ref{fig:example_flirting_preference} by the green transition arrow. This leads to $\omega_2$ upon which $m_1$ and $w_2$ block, yielding the matching $\mu_1$. This is followed by blocking from $m_2$ and $w_1$, yielding matching $\mu_2$, which is stable given $\omega_2$. This process is depicted in Figure~\ref{fig:example_flirting_preference_process}. Since $\omega_2$ is absorbing, we have reached a new self-confirming outcome. However, $(\mu_2, \omega_2$) can not be destabilized by further flirting. It is a ``flirt-proof'' self-confirming outcome while $(\mu_0, \omega_1$) is just a self-confirming outcome (that, as we have shown, is not flirt-proof). \hfill $\Box$\\

The example shows how flirting can lead to change of preferences through raising awareness and subsequently a blocking pair and common belief in blocking by this pair of players. A more subtle effect of flirting pertains just to the last feature. There are situations in which there is already a blocking pair but no common belief in blocking by this pair (e.g., Example 3). In such a case, flirting can help creating this pairwise common belief in blocking without any change of preferences. This is illustrated in the next example.\\
\begin{figure}[h]
    \centering
    \includegraphics[width=0.6\textwidth]{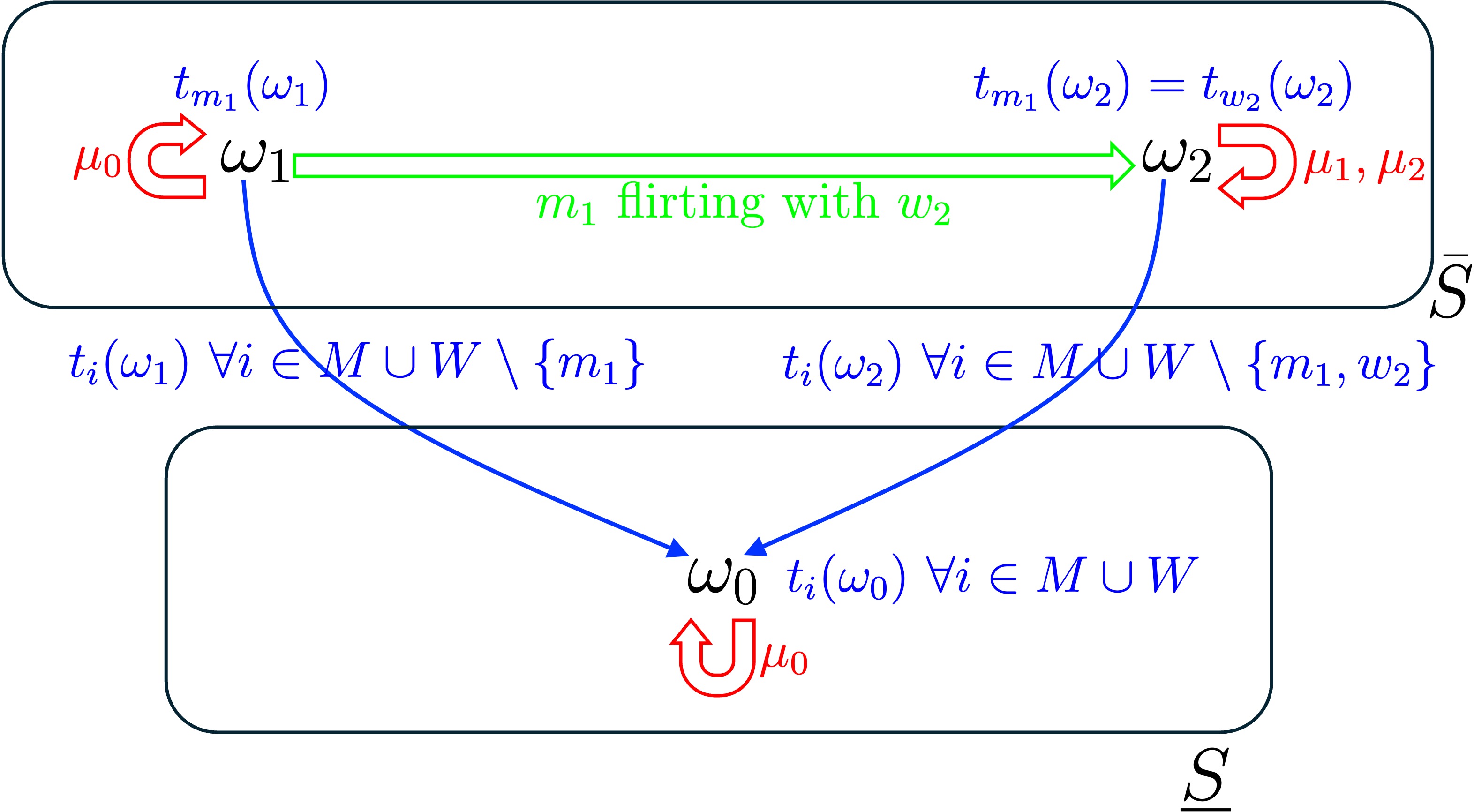}
    \caption{Unawareness Structure of Example 7}
    \label{fig:example_flirting_belief}
\end{figure}

\noindent \textbf{Example 7 } This example can be understood as an extension of Example 3. There are two men and women each, $M = \{m_1, m_2\}$ and $W = \{w_1, w_2\}$. The preference mappings are as before given by rank order lists of the prior Example 6. The unawareness structure is depicted in Figure~\ref{fig:example_flirting_belief}. Note that different from the prior example, at $\omega_1$ all players are unaware except man $m_1$. This is like in Example 3. Recall that $\mu_0$ is stable given $\omega_1$ in Example 3 because there is absence of pairwise common belief between $m_1$ and $w_2$ of blocking. However, at $\omega_1$, man $m_1$ could communicate with $w_2$ and raise her awareness, as indicated by the green state transition. At $\omega_2$ both are now aware and now $m_1$ and $w_2$ do not just form a blocking pair but there is also common belief among them of blocking. The resulting match $\mu_1$ leads to the final match $\mu_2$ at $\omega_2$, which is a self-confirming outcome. This outcome $(\mu_2, \omega_2$) can not be destabilized by further flirting. It is a ``flirt-proof'' self-confirming outcome while $(\mu_0, \omega_1$) is just a self-confirming outcome (that, as we have shown, is not flirt-proof). The process can be depicted as in Figure~\ref{fig:example_flirting_preference_process} except that now $m_1$ flirts with $w_2$ rather than the other way around. Thus, in this example flirting does not change preferences of the player who is flirted with but creates pairwise common belief of blocking (by making a player aware that \emph{others} have different preferences). \hfill $\Box$\\

We are interested in outcomes that are stable w.r.t. flirting, or as we alluded to already in the examples, are ``flirt-proof stable''. To define the refined stability notion, we need to model communication that raises awareness of potential blocking partners. This changes (point-)beliefs from one space to a richer space and thus consists of a transition to another state. We model this with an another finite state machine. In contrast to the transition function $\tau$ defined earlier, the transition due to communication will be part of the solution concept rather than the primitives of the dynamic matching game with unawareness. To this end, we require some notation. For any man $m \in M$, define $$H_m(\omega, \mu, w) := \{\omega' \in S_{t_m(\omega)}: \omega' \in CK_{m, w}([m \succ_w \mu(w)] \cap [w \succ_m \mu(m)]), S_{t_w(\omega')} \trianglerighteq S_{t_w(t_m(\omega))}\}$$ as the set of hypothetical states considered by $m$ at state $\omega$ in matching $\mu$ such that if $m$ would suitably raise $w$'s awareness, then there would be common belief in blocking among $w$ and $m$. Likewise, for any $w \in W$, define $H_w(\omega, \mu, m)$. 

Next, for any man $m \in M$, define $$S_m(\omega, \mu, w) := \bigvee_{\omega' \in H_m(\omega, \mu, w)} S_{t_w(\omega')}$$ as the awareness that is raised by $m$ to $w$ at state $\omega$ in matching $\mu$. Likewise, for any $w \in W$, define $S_w(\omega, \mu, m)$. Note that we assume that if there are alternative ways to raise awareness in order to obtain common belief in blocking, then the highest awareness is communicated.

Define the \emph{communication function} $f: \Omega \times \mathsf{M} \longrightarrow \Omega$ by for any $\omega \in \Omega$, $\mu \in \mathsf{M}$, we have that $f(\omega, \mu)$ with for any $w \in W$, $S_{t_w(f(\omega, \mu))} = S_{t_w(\omega)} \bigvee_{m \in M} S_m(\omega, \mu, w)$ and $m \in M$, $S_{t_m(f(\omega, \mu))} = S_{t_m(\omega)} \bigvee_{w \in M} S_w(\omega, \mu, m)$. By Assumption~\ref{ass: richness}, such a state $f(\omega, \mu)$ exists. 

After communication that potentially raises awareness of players and consequently change preferences, players may now want to communicate further. For any $(\omega, \mu)$ define recursively, $f^1(\omega, \mu) = f(\omega, \mu)$, and for $n > 1$, $f^n(\omega, \mu) = f(f^{n-1}(\omega, \mu), \mu)$ for any $\omega \in \Omega$ and $\mu \in \mathsf{M}$. That is, $f^n(\omega, \mu)$ captures $n$ rounds of communication starting from state $\omega$ and matching $\mu$. Since the model is finite and awareness can never decrease via communication, we observe: 

\begin{lem} For any $\omega \in \Omega$ and $\mu \in \mathsf{M}$, there exist a unique absorbing state for $f$ denoted by $f^{\infty}(\omega, \mu)$. 
\end{lem}

In the absorbing state of the communication function, further communication does not change awareness of any player. 

\begin{defin}[Flirt-proof stability] We say that matching $\mu$ is flirt-proof stable at $\omega$ if $\omega = f^{\infty}(\omega, \mu)$ and it is stable at $\omega$. 
\end{defin}

A flirt-proof stable matching is absorbing w.r.t. $f$ but not necessarily w.r.t. $\tau$. Adding latter, yields: 

\begin{defin}[Flirt-proof self-confirming outcome] We say that an outcome $(\omega, \mu)$ is \emph{flirt-proof self-confirming} if 
\begin{itemize} 
\item[(i)] $\mu$ is flirt-proof stable at $\omega$, and 
\item[(ii)] $\omega$ is absorbing (w.r.t. $\tau$) given $\mu$. 
\end{itemize} 
\end{defin} 

\begin{prop} Every finite dynamic two-sided matching game has a flirt-proof self-confirming outcome. 
\end{prop}

The proof is analogous to the proof of existence of a self-confirming outcome (Proposition~\ref{prop: existence}) and thus omitted. 

For a fixed small $\varepsilon \in (0, 1)$, define a matching and flirting process by transition probabilities $Q^{\varepsilon}$ such that for any $\omega$ and $\mu$, 
\begin{eqnarray*} Q^{\varepsilon}(\mu', \omega' \mid \mu, \omega) & := & \left\{ \begin{array}{cl} \frac{1 - \varepsilon}{|\widehat{MB}(f^{\infty}(\omega, \mu), \mu)|}  & \mbox{if } \mu' \mbox{ differs from } \mu \mbox{ by satisfaction of exactly one} \\ & \mbox{pair in } \widehat{MB}(f^{\infty}(\omega, \mu), \mu) \mbox{ and } \tau(f^{\infty}(\omega, \mu), \mu') = \omega'; \\ \\
\frac{\varepsilon}{|B(f^{\infty}(\omega, \mu), \mu)|} & \mbox{if } \mu' \mbox{ differs from } \mu \mbox{ by satisfaction of exactly one} \\ & \mbox{pair in } B(f^{\infty}(\omega, \mu), \mu) \mbox{ and } \tau(f^{\infty}(\omega, \mu), \mu') = \omega'; \\ \\
1 & \mbox{if }  \mu' = \mu \mbox{ and } \mu \mbox{ is flirt-proof stable given } \omega \mbox{ and} \\
& \tau(f^{\infty}(\omega, \mu), \mu') = \omega'; \\  \\
0 & \mbox{else. }
\end{array} \right.  
\end{eqnarray*} 

Naturally, the process would proceed as follows: Given a current state $\omega$ and awareness/preferences at $\omega$, the current matching $\mu$, some players may have incentives to raise awareness of others, and the state evolves through communication to $f(\omega, \mu)$ and further until communication quiets down and reaches state $f^{\infty}(\omega, \mu)$. Now there may be some blocking pairs w.r.t. preferences at $f^{\infty}(\omega, \mu)$. Satisfying some blocking pair, with $(1 - \varepsilon$)-priority given to a mutually optimal blocking pair, would lead to another matching $\mu'$. At this matching $\mu'$ and state $f^{\infty}(\omega, \mu)$, there may be discoveries leading to another state $\tau(f^{\infty}(\omega, \mu), \mu') = \omega'$. At this state and corresponding awareness/preferences, some players may have an urge to communicate and raise other players' awareness. This goes on until communication quiets down again and there might be now some pairwise common belief of blocking. Satisfying such a blocking pair, with $(1 - \varepsilon)$-priority given to a mutual optimal best blocking pair, may lead to yet another match $\mu''$ ... We let states transit \emph{after} the transition of the matching, modeling transition through transformative experiences in the matching. We also let states transit \emph{before} the transition of the matching, modeling transition through communication. Above formal description of the process involves a simultaneous transition of states and matches. It can be thought of just describing the process at every ``even'' period. In a sense, the implicit ``odd'' period consists of multiple rounds of communication between players and the transition of the matching and the implicit ``even'' period consists of the transformative experiences. 

\begin{lem} An outcome $(\mu, \omega)$ is flirt-proof self-confirming if and only if it is an absorbing outcome of $Q^{\varepsilon}$. 
\end{lem} 

\noindent \textsc{Proof. } $\Leftarrow:$ Let $(\mu, \omega)$ be an absorbing outcome of $Q^{\varepsilon}$. Then $Q^{\varepsilon}(\mu, \omega \mid \mu, \omega) = 1$. By the construction of $Q^{\varepsilon}$,  $(\mu, \omega)$ is flirt-proof self-confirming. 

$\Rightarrow:$ Let $(\mu, \omega)$ be flirt-proof self-confirming. Then $\mu$ is flirt-proof stable at $\omega$, and  $\omega$ is absorbing given $\mu$.  Hence, $Q^{\varepsilon}(\mu, \omega \mid \mu, \omega) = 1$, i.e., $(\mu, \omega)$ be an absorbing outcome of $Q^{\varepsilon}$.\hfill $\Box$

\begin{prop} For any (initial) outcome $(\mu, \omega)$ and $\varepsilon \in (0, 1)$, the matching and flirting process $Q^{\varepsilon}$ convergences to a flirt-proof self-confirming outcome almost surely. 
\end{prop}

\noindent \textsc{Proof. } First, we argue that each absorbing set of $Q^{\varepsilon}$ cannot involve more than one state (denoted with $\omega^*$). Suppose by contradiction that there exists an absorbing set of $Q^{\varepsilon}$ that involves two different states. Denote them by $\omega^*$ and $\omega^{**}$ with $\omega^* \neq \omega^{**}$. By the definition of absorbing set, each outcome (consisting both of a state and a matching) is reachable from any other outcomes in the absorbing set via a finite number of transitions. Thus, $\omega^*$ must be reachable from $\omega^{**}$ and vice versa. That is, $\tau(...\tau(f^{\infty}(\omega^*, \cdot), \cdot) ...,\cdot) = \omega^{**}$ and $\tau(...\tau(f^{\infty}(\omega^{**}, \cdot), \cdot) ...,\cdot) = \omega^*$. Since for each agent $i$, awareness can only increase along the process, we must have $S_{t_i(\omega^*)} = S_{t_i(\omega^{**})}$ for all $i$. The assumption of no redundancies, Assumption~\ref{ass: redundancies}, implies now that $\omega^* = \omega^{**}$, a contradiction. 

The rest of the proof is analogous to the proof of Proposition~\ref{prop: convergence}. \hfill $\Box$\\ 

Recall that $(\mu, \omega)$ is self-confirming if (i) $\mu$ is stable at $\omega$, and (ii) $\omega$ is absorbing given $\mu$, i.e., $\tau(\omega)=\omega$. Meanwhile, $(\mu, \omega)$ is flirt-proof self-confirming if (i) matching $\mu$ is stable at $\omega$, (iii) $\omega = f^{\infty}(\omega, \mu)$, and (iv) $\tau(\omega) = \omega$ absorbing given $\mu$. Since (i)-(iii) implies (i)-(ii), flirt-proof self-confirming implies self-confirming. Thus, we conclude: 

\begin{prop} If an outcome $(\mu, \omega)$ is flirt-proof self-confirming, then it is also a self-confirming outcome.  
\end{prop}

Examples 6 and 7 show that the converse does not hold.

\section{Does Divorce Improve Welfare?\label{sec:divorce}}

Will divorcers, who experienced preference changes, necessarily become better off? More precisely, suppose a player is enlightened during a matching, changes his/her preferences, and consequently divorces his/her current match. Will such a player become necessarily better off in the resulting rematching process where the player's welfare is evaluated using his/her new preferences? Similarly, will the divorcee, i.e., the player who is divorced by the divorcer, necessarily become worse off? We show by example that this is not the case. In terms of welfare of the divorcer and the divorcee, anything goes even within the same matching game and the same initial condition.\\

\noindent \textbf{Example 8 } Suppose there are five men $M = \{m_1, m_2, m_3, m_4, m_5\}$ and five women $W = \{w_1, w_2, w_3, w_4, w_5\}$. Consider the following strict preferences given by the rank order lists: 
$$\begin{array}{ll} \succ_{m_1}: & w_1, w_2, w_4, w_3, w_5 \\
\succ_{m_2}: & w_2, w_3, w_4, w_1, w_5 \\
\succ_{m_3}: & w_1, w_3, w_2, w_4, w_5 \\
\succ_{m_4}: & w_4, w_5, w_3, w_2, w_1 \\
\succ_{m_5}: & w_5, w_1, w_3, w_2, w_4 \\
\succ_{w_1}: & m_5, m_1, m_3, m_4, m_2 \\
\succ_{w_2}: & m_3, m_1, m_2, m_5, m_4 \\
\succ_{w_3}: & m_2, m_3, m_1, m_4, m_5 \\
\succ_{w_4}: & m_1, m_2, m_4, m_3, m_5 \\
\succ_{w_5}: & m_4, m_5, m_1, m_2, m_3 
\end{array}$$
and the initial stable matching $\mu_1 = \begin{pmatrix}
m_1 & m_2  & m_3  & m_4 & m_5\\
w_1 & w_2 & w_3 & w_4 & w_5
\end{pmatrix}$. 
\begin{figure}[h]
    \centering
    \includegraphics[width=1\textwidth]{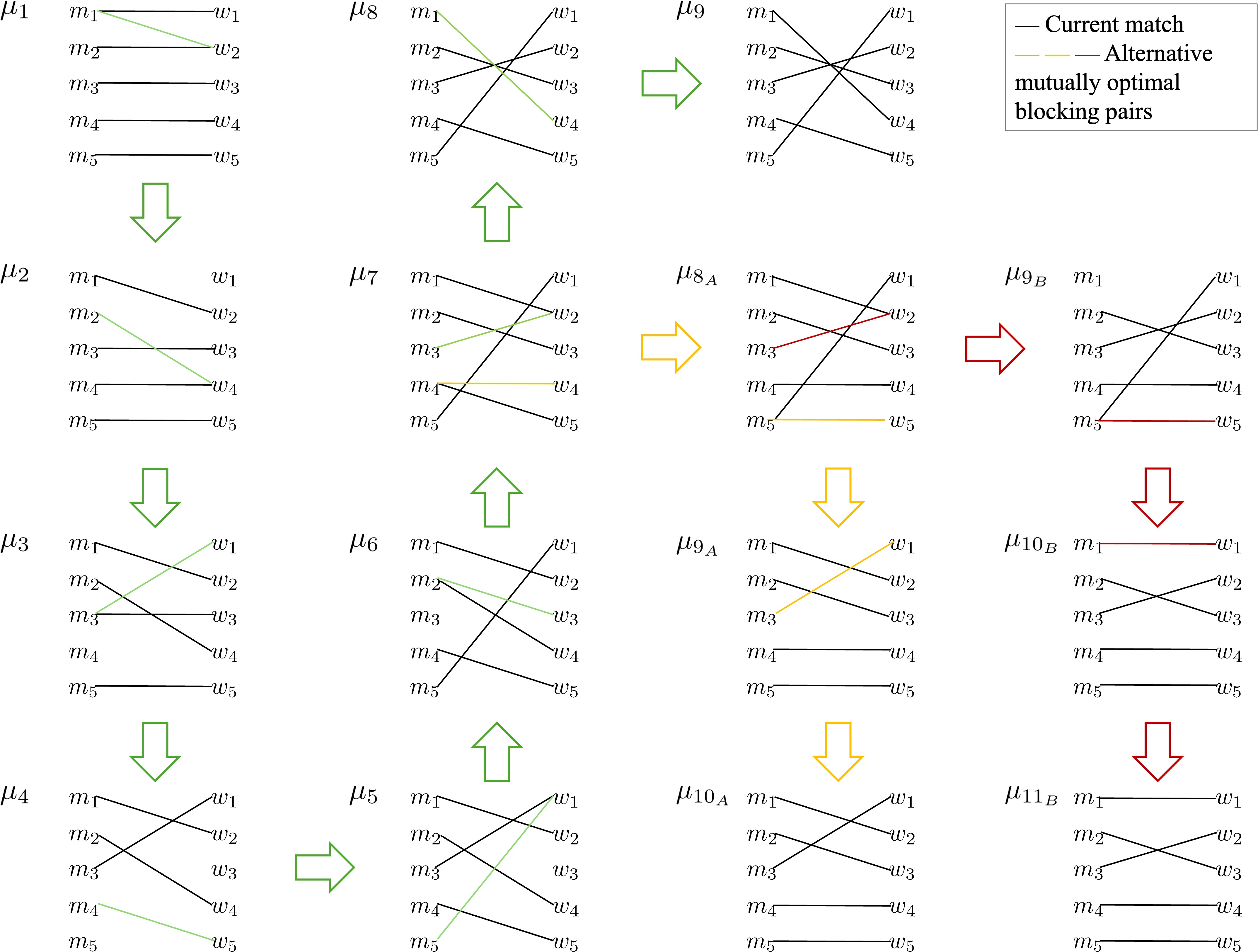}
    \caption{Processes in Example 8\label{fig:welfare}}
\end{figure}

Suppose $m_1$ is enlightened and changes his preference such that his new rank order list is $w_2, w_1, w_4, w_3, w_5$. For simplicity, suppose that this is the only change in the players' preferences along the entire process and that all players have correct beliefs about every other agents' preferences. With $m_1$'s new preference, he blocks with $w_2$. We show that if $m_1$ rematches to $w_2$, the process of satisfying mutually optimal blocking pairs may lead to any of the three stable matchings, depending on which mutually optimal blocking pair is satisfied at stages 7 or 8 (see Figure~\ref{fig:welfare}). Furthermore, in these stable matchings, $m_1$ can be matched to $w_2$ (in $\mu_{10_A}$), $w_1$  (in $\mu_{11_B}$), or $w_4$  (in $\mu_{9}$), which means that he may be better off, same, or worse off. 

Now observe woman $w_1$. She is the partner that gets initially divorced by the divorcer $m_1$. After the rematching process, she may be matched to $m_5$ (in $\mu_9$, her most preferred counterpart, making her strictly better off compared to the initial matching. She may also be matched to $m_3$ (in $\mu_{10A}$), making her worse off compared to her initial match. Finally, she may also be matched to $m_1$ again (in $\mu_{11B}$), her initial match, resulting in no change of welfare for her. Thus, the example demonstrates that for both the divorcer and divorced may be better, worse, or equal as well off as in the initial matching, even within the same matching game and from the same initial condition.\hfill $\Box$

\section{Discussion\label{sec:discussion}} 

\subsection{Infidelity\label{sec:infidelity}} 

\begin{figure}[h]
    \centering
    \includegraphics[width=0.8\textwidth]{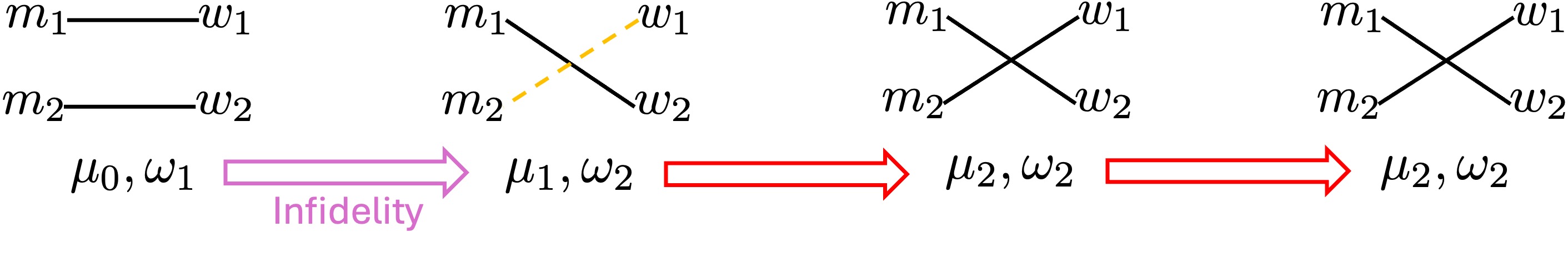}
    \caption{Infidelity in Example 5}
    \label{fig:example_trap_matching_infidelity}
\end{figure}

Neither in self-confirming outcomes nor flirt-proof self-confirming outcomes, players are guaranteed to become fully aware. The reason is that stability given the state prevents them from making transformative experiences and absorbing states does not allow for further changes of preferences in their stable matching. This has been illustrated in prior Example 5. One extra source of experiences is experimentation. In the marriage market it may be dubbed ``infidelity''. Consider again Example 5. Suppose that in the self-confirming outcome $(\mu_0, \omega_1)$ both man $m_1$ and $w_2$ temporary match despite $\mu_0$ being stable given $\omega_1$. (Recall that in matching $\mu_0$ given by the first matching in Figure~\ref{fig:example_trap_matching_infidelity}, man $m_1$ is matched to woman $w_1$ and woman $w_2$ is matched to man $m_2$.)  Then man $m_1$ and $w_2$ would become aware that they are each others' first choice. The state would transit to $\omega_2$. Consequently, $m_1$ and $w_2$ form a blocking pair and since this is also common belief among $m_1$ and $w_2$, the original matching $\mu_0$ would be destabilized. Divorcing the players and satisfying the blocking pair yields matching $\mu_1$ at state $\omega_2$; see Figure~\ref{fig:example_trap_matching_infidelity}. This may be dubbed the ``direct'' effect of infidelity. Yet, there is also an indirect effect. Matching $\mu_1$ is not stable given $\omega_2$ because both man $m_2$ and woman $w_1$ are unmatched and now form a blocking pair and this is common belief. Consequently, satisfying this blocking pair yields matching $\mu_2$; see Figure~\ref{fig:example_trap_matching_infidelity} for the resulting sequence of matchings. In matching $\mu_2$, both man $m_2$ and woman $w_2$ would become also aware that they are each others' first choice. This is the ``indirect'' effect of infidelity of $m_1$ and $w_2$ on $m_2$ and $w_2$. Contrary to the typical view of infidelity, it is a positive external effect in this case. (Obviously, this is not generally the case.) We reach a self-confirming outcome $(\mu_2, \omega_2)$, in which all parties are fully aware. In this example, infidelity allows all to escape the unawareness trap by marriage. 

\subsection{Stable Confusion About Others\label{sec:confusion}} 

Consider again Example 7. In outcome $(\mu_2, \omega_2)$, man $m_1$ is matched to woman $w_2$ and man $m_2$ is matched to woman $w_1$. While it is pairwise common belief among $m_1$ and $w_2$ that they prefer each other over the other counterpart, it is not common belief among all players. In particular, since at state $\omega_2$ the point-belief of both man $m_1$ and woman $w_2$ is $\omega_0$ (see Figure~\ref{fig:example_flirting_belief}), they do not understand why $m_1$ and $w_2$ formed a blocking pair and why the matching changed from $\mu_0$ to $\mu_2$. In other words, they are confused about $m_1$ and $w_2$. Note also that players $m_1$ and $w_2$ can not gain anything from enlightening players $m_2$ and $w_2$. So there is no force, either by communication or blocking actions, to change the situation. The outcome is flirt-proof self-confirming. It emphasizes the fact that a matching can be stable given beliefs and beliefs can be stable given the matching despite some players being confused. The confusion itself is ``stable'' because there is nothing they can do about within the game. We believe that this is often quite realistic in matching markets. It is also not surprising from a theoretical point of view because solution concepts to coalitional games like stability or the core are not concepts akin to rationalizability in non-cooperative game theory. 

While self-confirming outcomes do not rationalize the confusion, we may wonder about the possible state of mind of players who are confused. There are two possible ways to rationalize such confusion. The first is that others make mistakes: They should block but mistakenly do not do so. The second explanation keeps the assumption that others are rational but explains the confusion with awareness of unawareness: When a player expects others to block but such blocking does not happen, the player may suspect that (e)he her/himself is unaware of something that some others are aware. That is, the player may become aware that (s)he is unaware of something. While our model can be extended to model awareness of unawareness explicitly using the tools presented in Schipper (2024), it would not change any of our conclusions unless players can do something about discovering what they might miss. This would require enriching the problem with additional structure such as infidelity as discussed above or individual actions of engaging in gossiping, asking for counseling, etc.

\subsection{Related Literature\label{sec:related_literature}} 

Our work is related to five strands of literature. The first strand is the literature on matching with incomplete information. There are several papers on matching with incomplete information under the non-transferable utility (NTU) framework. Under two-sided uncertainty, Lazarova and Dimitrov (2017) study stability where a pair of agents blocks if each of the agents believe that he/she can do better with positive probability. This is a very demanding notion of stability as a stable matching must be robust to beliefs that put a tiny probability on agents do better. Our notion of stability, in contrast, is absence of pairwise common (point-)belief in blocking. In order to block a pair must \emph{agree} to block, i.e., it must be common belief among them that both prefer each other over their current partners. Under one-sided uncertainty where the workers' types distribution and the firms' types are common knowledge, Bikhchandani (2017) investigates two notions of stability: Ex ante stability where agents block if they are better off with all admissible types and Bayesian stability where agents block if they have higher expected utilities. Ex ante stability implies that participation in a block makes it common certainty that each is better off in a block. The notion of ex ante stability is adapted from a notion of stability introduced for transferable utility (TU) matching games with one-sided incomplete information by Liu et al. (2014). Pomatto (2022) provides an epistemic non-cooperative counterpart to Liu et al. (2014). Chen and Hu (2020) construct a learning process leading to an extension of the notion of stability by Liu et al. (2014). Alston (2020) shows that belief restrictions imposed on top of the stability notion by Liu et al. (2014) may lead to non-existence (see also Bikhchandani, 2017). Liu (2020) offers a stability notion for TU matching games with one-sided incomplete information that is explicit about the on-path and off-path beliefs. The stability notion closest to ours in the literature on TU matching with incomplete information is Forges (2004), who studies the extension of the incentive compatible coarse core of Vohra (1999) to assignment games. There is also the literature on centralized matching with incomplete information from a mechanism design perspective; see for instance Roth (1989), Majumdar (2003), Ehlers and Mass\'{o} (2007), Yenmez (2013), and Fernandez, Rudov, and Yariv (2022). 

The second strand is the literature on decentralized matching. Roth and Vande Vate (1990) showed that from any matching, there exists a sequence of matching by satisfying blocking pairs that leads to a stable matching. Ackermann et al. (2008) showed that from any matching, there exists a sequence of matching by satisfying optimal blocking pairs that leads to a stable matching. As corollaries, the process of satisfying random (optimal) blocking pairs leads to a stable matching with probability one.\footnote{Applications of Roth and Vande Vate (1990) on random paths to stability include decentralized market processes for stable job matching with competitive salaries (Chen, Fujishige, and Yang, 2011) and paths to stability under incomplete information (Lazarova and Dimitrov, 2017, for under NTU matching games and Chen and Hu, 2020, for TU matching games). It has also been extended to more general contexts such as matching with couples (Klaus and Klijn, 2007), many-to-many matching (Kojima and  {\"U}nver, 2008), and many-to-many matching with contracts (Mill{\'a}n and Pepa Risma, 2018).} These results are partly driving our decentralized matching process to converge to a self-confirming outcome: As we perturb the process that prioritizes mutually optimal blocking pairs with $\varepsilon$ probability to select an optimal but not necessarily mutually optimal blocking pair even when a mutually blocking pair exists, we always have positive probability over all optimal blocking pairs. This is very different from Klaus, Klijn, and Walzl (2010) who perturb their process with $\varepsilon$ probability to match a non-blocking pair together. Another difference between our random path to self-confirming outcomes and random path to stability in the literature is that our notion of blocking differs slightly since we require pairwise common (point-)belief in blocking. An insightful paper on random path to stability is Rudov (2024), who shows that under some conditions, any unstable matching can reach any stable matching through the process of satisfying random (optimal) blocking pairs. As mentioned already in Section~\ref{section:Zhang}, he also observed using a five-by-five market that a process that satisfies only mutual optimal blocking pairs whenever the exist may fail to converge to a stable matching. From non-cooperative or search perspectives, Lauermann and N\"{o}ldeke (2014), Wu (2015), and Ferdowsian, Niederle, and Yariv (2025) show that decentralized interactions only lead to stable outcomes when there is a unique stable matching or when preferences are highly correlated. In this context note that our observation in Section~\ref{sec: revisited} that without restrictions on the marriage market structure, prioritizing mutually optimal blocking pairs can lead to cycles, is consistent with these findings. Adachi (2003) shows that equilibrium outcomes converge to stable matchings as search frictions vanish. Search frictions in Adachi (2003) are represented by a time discounting while we interpret frictions is any factor that favors a non-mutual optimal blocking pair over a mutually optimal blocking pair. In contrast to Adachi (2023), we show that frictions can be arbitrary small but should not vanish for convergence to a stable matching. Doval (2022) investigates a notion of dynamic stability when matching opportunities arrive over time and matching is irreversible. In an interesting paper combining both incomplete information and decentralized non-cooperative matching in a labor market context, Ferdowsian (2024) allows workers to learn about their liking of a firm via getting matched to the firm. Matchings are broken after each period and learning is immediate upon a match. Besides the different framework and setting, a main difference to our paper is that all learning is anticipated. 

The third strand is the literature on the core. When modeling a matching problem under complete information, the core of the marriage market is the set of stable matchings (Roth and Sotomayor, 1990, Sasaki and Toda, 1992). For exchange economies with asymmetric information, Wilson (1978) introduced the notions of the coarse core and the fine core, which involves no information sharing or maximal information sharing respectively. Since both the core allocations and counterfactual blocking may be informative, these core concepts have been further refined (e.g., Vohra, 1999, Forges, 2004, Ray and Vohra, 2015). For surveys of the core under incomplete information, see Forges, Minelli, and Vohra (2002) and Forges and Serrano (2013). The coarse core has been extended to coalitional games with unawareness by Bryan, Ryall, and Schipper (2022). Our notion of stability is inspired by the coarse core. Our notion of flirt-proof stability is inspired by refinements of the coarse core that feature intentional communication. In some sense, flirt-proof stability combines the idea of the coarse core and with the idea of non-cooperative disclosure games (e.g., Milgrom and Roberts, 1986) but keeps the spirit of cooperative game theory.

The fourth strand is the literature on communication and disclosure in matching games with incomplete information. While Hoppe, Moldovanu, and Sela (2009) and Coles, Kushnier, and Niederle (2013) study costly signaling in matching games, Ostrowsky and Schwarz (2010), Bilancini and Bonicelli (2013), and Chade and Pram (2024) study (costly) disclosure of information in matching games. All these papers make use of ideas from non-cooperative game theory for equilibria under communication while arguably we model raising awareness via flirting in the spirit of cooperative game theory. Moreover, none of these papers consider raising awareness. 

The final strand is the literature on unawareness. Since this is the first paper on matching under unawareness, we contribute to the recent growing literature on exploring the implications of unawareness in economics. Other applications of unawareness pertain to disclosure, moral hazard, contract theory, screening, efficient mechanism design, auctions, procurement, delegation, speculation, financial market microstructure, default in general equilibrium, electoral campaigning, business strategy, and conflict resolution; for a bibliography, see Schipper (2025).

\end{document}